\documentclass[prl,twocolumn,superscriptaddress]{revtex4-1}
\usepackage[dvipdfmx]{graphicx}
\usepackage{graphicx}
\usepackage{physics}
\usepackage{bm, amsmath, amssymb, braket}
\usepackage{times}
\usepackage{multirow}
\usepackage{ascmac}
\usepackage{amsthm}
\usepackage{float}
\usepackage{color}
\usepackage{comment}
\usepackage{mathtools}  % for math
\usepackage{mathrsfs} 
\usepackage[colorlinks=True,urlcolor=blue,linkcolor=blue,citecolor=blue]{hyperref}% add hypertext capabilities
\usepackage{xcolor}
\usepackage{multirow}
\usepackage{setspace}
\usepackage{subfigure}

\usepackage[toc,page]{appendix}

\begin{document}
\title{Probing quantum critical crossover via impurity renormalization group}
\author{Tao Yang}
\affiliation{National Laboratory of Solid State Microstructures and Department of Physics, Nanjing University, Nanjing 210093, China}

\author{Z. Y. Xie}
\email{qingtaoxie@ruc.edu.cn}
\affiliation{School of Physics, Renmin University of China, Beijing 100872, China}
\affiliation{Key Laboratory of Quantum State Construction and Manipulation (Ministry of Education), Renmin University of China, Beijing 100872, China}

\author{Rui Wang}
\email{rwang89@nju.edu.cn}
\affiliation{National Laboratory of Solid State Microstructures and Department of Physics, Nanjing University, Nanjing 210093, China}
\affiliation{Collaborative Innovation Center of Advanced Microstructures, Nanjing University, Nanjing 210093, China}
\affiliation{Jiangsu Physical Science Research Center}
\affiliation{Hefei National Laboratory, Hefei 230088, People's Republic of China }

\author{Baigeng Wang}
\affiliation{National Laboratory of Solid State Microstructures and Department of Physics, Nanjing University, Nanjing 210093, China}
\affiliation{Collaborative Innovation Center of Advanced Microstructures, Nanjing University, Nanjing 210093, China}
\affiliation{Jiangsu Physical Science Research Center}

\begin{abstract}
Quantum impurities can host exotic many-body states that serve as sensitive probes of bath correlations. However, quantitative and non-perturbative methods for determining impurity thermodynamics in such settings remain scarce. Here, we introduce an impurity renormalization group approach that merges the tensor-network representation with the numerical renormalization group cutoff scheme. This method overcomes conventional limitations by treating bath correlations and impurity interactions on an equal footing. Applying our approach to the finite-temperature quantum critical regime of quantum spin systems, we uncover striking impurity-induced phenomena. In a coupled Heisenberg ladder, the impurity triggers a fractionalization of the local magnetic moment. Moreover, the derivative of the impurity susceptibility develops cusps that mark the crossover into the quantum critical regime. We also observe an exotic evolution of the spin correlation function driven by the interplay between bath correlations and the impurity. Our results demonstrate that this method can efficiently solve correlated systems with defects, opening new pathways to discovering novel impurity physics beyond those in non-interacting thermal baths.

%We further apply the approach to study the impurity effects on the finite-temperature quantum critical regime of quantum spin systems. Remarkably, we show that the impurity induces a fractionalization phenomena of local momentum in the coupled Heisenberg ladder model. Moreover, it generates cusps in the derivative of the impurity susceptibility, which acts as an indicator of the crossovers associated with the quantum critical regime. An exotic evolution behavior of the spin correlation function is also revealed, as a combined effect of the bath correlation and the impurity. These results justify that our developed approach could efficiently solve correlation problems intertwined with defects, opening new possibilities to uncover novel impurity physics absent in non-interacting thermal baths. We introduce an impurity renormalization group approach for thermodynamic systems, which merges the representation framework of the tensor-network renormalization group with the cutoff scheme of the numerical renormalization group. This framework overcomes conventional limitations by treating bath correlations and impurity interactions on an equal footing.  that applicable to systems in thermodynamic limit
\end{abstract}
\maketitle
\emph{\color{blue}{Introduction.--}} 
%Quantum impurity problems, describing localized quantum states coupled to a large reservoir, play a profoundly important role in condensed matter physics. Intensive studies in this field not only brought about fundamental concepts such as the Kondo effect [Cite] and the non-Fermi liquid physics [Cite], but also stimulated development of powerful numerical approaches to strongly-correlated systems, including the numerical renormalization group (NRG) [Cite Wilson, NRG review] and the dynamical mean-field theory [Cite Gabby]. More importantly, the impurity or defect offers an efficient quantum probe of essential physical properties of the underlying reservoir. For example, it can give accurate hints about the host superconductors, not only in terms of their pairing symmetry [Cite JXZ] but also their internal symmetries [Cite RW], etc. Hence, solving the quantum impurity problem advances the understanding of the bath, promising a general quantum probe of exotic quantum phenomena in many-body systems.
Quantum impurity problems, which describe localized quantum degrees of freedom coupled to an extended reservoir, are of profound importance in condensed matter physics. Intensive research in this area has not only led to fundamental conceptual breakthroughs—such as the Kondo effect \cite{Kondo1964, nozieres1980, andrei1983, hewson1997kondo, goldhaber1998} and non-Fermi liquid behavior \cite{Affleck1991, Kevin1992, Cox1993, Cox1996, GMZhang1996, lee2018recent}—but has also driven the development of powerful numerical methods for strongly correlated systems, including the numerical renormalization group (NRG) \cite{Wilson1975, bulla2008numerical}, the dynamical mean-field theory (DMFT) \cite{DMFT, kotliar2006}, and other more recent proposals \cite{wang2010impurity, Lu2015, Lu2023}. Moreover, impurities and defects serve as effective quantum probes, capable of revealing essential physical properties of the host system. For instance, they can provide precise insights into the pairing symmetry \cite{JXZ, zhang2009} and internal symmetries \cite{RW2019} of various superconductors. As such, solving quantum impurity problems establishes a versatile framework for probing exotic quantum phenomena in the host materials \cite{XQW, XCX2023}.

%When the host materials are strongly-correlated in nature, the local impurity could generate more
%exotic correlated physics, as studied in spin liquids [Cite RW, Kitaev] and high $T_c$ superconductors [Cite]. However, unlike the non-interacting reservoirs \cite{note3}, the accurate method that solves impurity problems in strongly-interacting baths is yet to be established. As shown in Fig.~\ref{fig1}(a), for reservoir with vanishing interaction $V_{ij}=0$, the standard NRG impurity-solver maps the problem into a half-infinite Wilson chain. Then, the iterative diagonalization of the chain generates the impurity thermodynamics, which usually saturates at a cutoff in terms of the chain length, i.e., $\Lambda_{\mathrm{imp}}$. This defines a scale up to which the impurity effect is extended into the reservoir, separating an impurity and a bulk region as indicated by Fig.~\ref{fig2}(b). In sharp contrast, for strongly-correlated reservoir with $V_{ij}\neq0$, such a mapping scheme is in principle inapplicable, and the bath correlation must be treated at an equal footing with the impurity-bath coupling. This poses a fundamental challenge that hinders the accurate solution of the impurity state. 

\begin{figure}
\includegraphics[width=\linewidth]{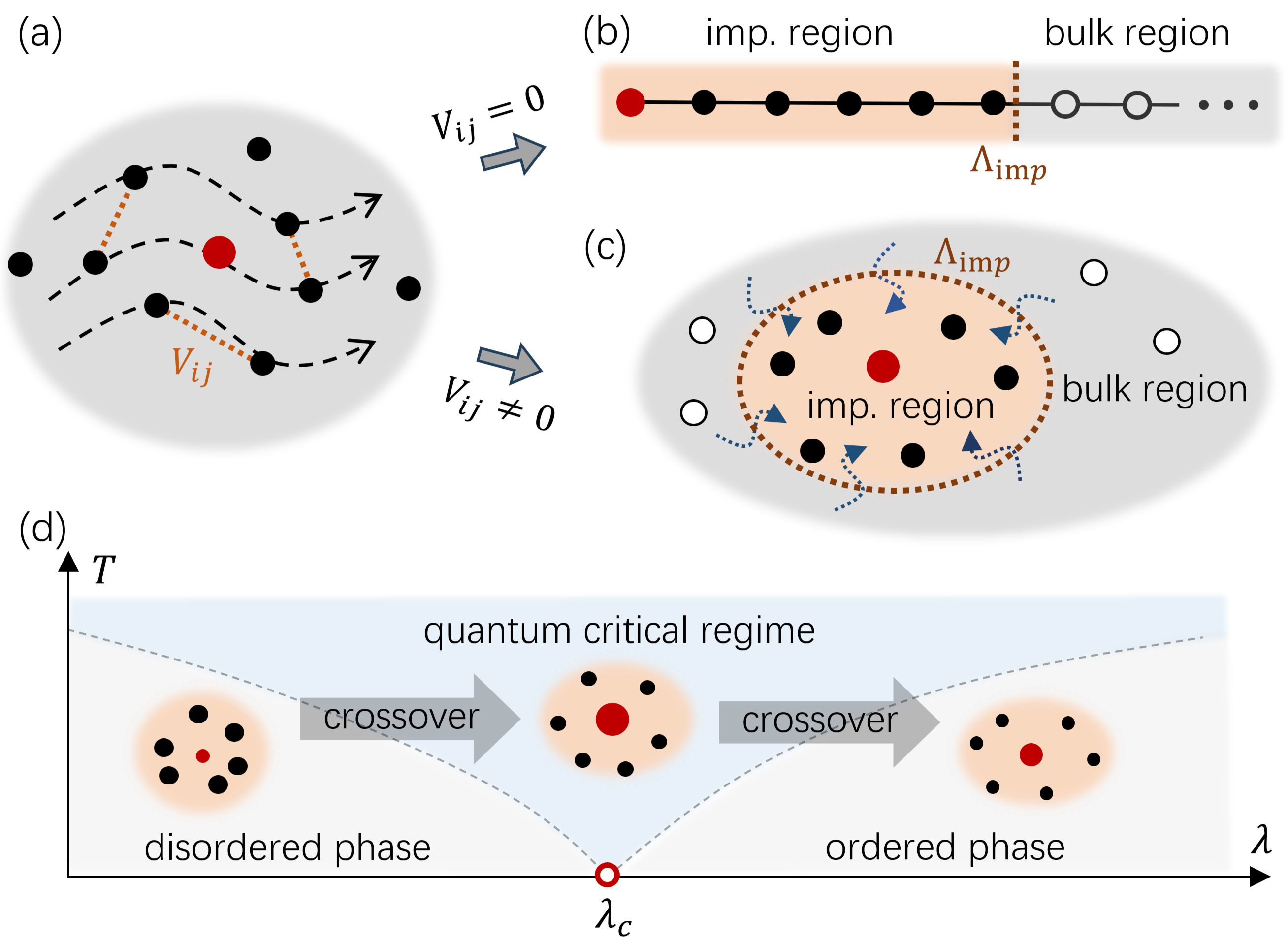}
\caption{\label{fig1} Impurity problems in non-interacting baths can be mapped to a Wilson chain with a length cutoff $\Lambda_{\mathrm{imp}}$ ((a) and (b)). For a strongly-correlated bath, the chain mapping scheme is not applicable, but the impurity problems can be represented by local tensors defined in an impurity and a bulk region, separated by the cutoff $\Lambda_{\mathrm{imp}}$ ((a) and (c)). (d) The impurity can be used as a local quantum probe to diagnose the QCR and crossovers associated with quantum phase transitions. }
\end{figure}

When the host material is strongly correlated, a local impurity can give rise to more exotic correlated phenomena, as studied in spin liquids \cite{RW2021, Willans2010, Takahashi2023, Yatsuta2024, Onur} and high-$T_c$ superconductors \cite{vojta2000, hudson2001, JXZ2001, balatsky2006}. However, in contrast to non-interacting reservoirs \cite{note3}, a general and accurate method for solving impurity problems embedded in strongly interacting baths has not yet been established. As depicted in Fig. \ref{fig1}(a), for a reservoir with vanishing interaction ($V_{ij}=0$), the standard NRG impurity solver maps the problem onto a semi-infinite Wilson chain. Iterative diagonalization of the chain then yields the impurity thermodynamics, which typically saturate at a cutoff chain length, denoted as $\Lambda_{\mathrm{imp}}$. This scale defines the extent to which the impurity affects the reservoir, effectively dividing the system into an impurity region and a bulk region, as illustrated in Fig. \ref{fig1}(b). In stark contrast, for a strongly correlated reservoir with $V_{ij}\neq0$, such a mapping scheme is in principle not applicable, as the bath correlations must be treated on an equal footing with the impurity-bath coupling. This poses a fundamental challenge that hinders an accurate solution for the impurity state.

%In this Letter, we develop a controllable impurity renormalization group (IRG) approach that accurately solves the impurity problems in correlated thermal bath. In stark contrast to the Wilson chain scheme in Fig.~\ref{fig2}(b), we show that the problem can be naturally represented by two local tensors defined in the impurity and the bulk region, respectively (Fig.~\ref{fig1}(c)). Then, we propose a finite-temperature approach that produces all the impurity thermodynamics based on the tensor-network renormalization group [Cite ZYX, WL].  The bulk correlation and the impurity coupling are also treated at an equal footing in the thermodynamic limit, which well overcomes the long-standing challenge in the impurity physics.

In this Letter, we develop a controllable impurity renormalization group (IRG) approach to accurately solve impurity problems within correlated thermal baths. We demonstrate that the problem admits a natural representation via two local tensors, defined respectively in the impurity and bulk regions (Fig.~\ref{fig1}(c)), similar to the cutoff scheme depicted in Fig.~\ref{fig1}(b). We then introduce a finite-temperature framework, built upon the tensor-network renormalization group \cite{Jiang2008, Verstraete2008, Gu2008, srg, HHZhao2010, WeiLi2011, WeiLi2012, PESS, WeiLi2018, ORUS2014, orus2019tensor, Orus2019finiteT}, which provides a complete account of the impurity thermodynamics. Crucially, our method treats bulk correlations and impurity coupling on an equal footing directly in the thermodynamic limit, thereby overcoming a long-standing challenge in impurity physics.

Using our developed IRG method, we investigate quantum critical phenomena at finite temperatures. A zero-temperature quantum critical point generically induces a quantum critical regime (QCR) with crossovers to neighboring regimes, as illustrated in the $T$-$\lambda$ phase diagram of Fig. \ref{fig1}(d), where $\lambda$ is the parameter driving the quantum phase transition. Within the QCR, physical observables are expected to follow a power-law dependence on $T$ and a smooth dependence on $\lambda$, governed by a universal scaling function. While the temperature dependence has been extensively studied, the $\lambda$-evolution at fixed $T$ remains much less explored. To this end, we study a defect embedded in a coupled spin-ladder system hosting a valence-bond-solid (VBS) to antiferromagnetic (AFM) phase transition. By monitoring the evolution of the impurity states with tuning $\lambda$, we probe key properties of the QCR and its crossovers (Fig. \ref{fig1}(d)). Remarkably, the IRG method uncovers the emergence of a fractionalized local moment near the defect \cite{sachdev1999}. Furthermore, we observe ``cusps" in the derivative of the impurity susceptibility, which serve as a numerical indicator for the crossover towards the QCR. We also identify a columnar to pinwheel-like evolution in the spin-spin correlation texture with increasing $\lambda$. These findings reveal new aspects of the QCR previously inaccessible by established methods, thereby demonstrating the power and efficiency of our IRG approach.

%In conventional approaches t quantum impurity models, such as the numerical renormalization group (NRG) method, it is common to integrate out the bath degrees of freedom. This results in an effective impurity model that is further mapped to an 1D chain. Such a scheme applies to cases where the baths are non-interacting or weakly interacting with Fermi liquid type descriptions. However, for cases where the interactions within the baths are non-negligible, the baths can not be treated as Gaussian environments for the impurity. Consequently, the above approach does not apply, and the bath interaction must be treated at an equal footing with the impurity-bath interaction. Thus,  in order to accurately obtain thermal dynamic properties for quantum impurities in strongly-interacting baths,  controllable numerical methods are yet to be established.

\begin{figure}
\includegraphics[width=\linewidth]{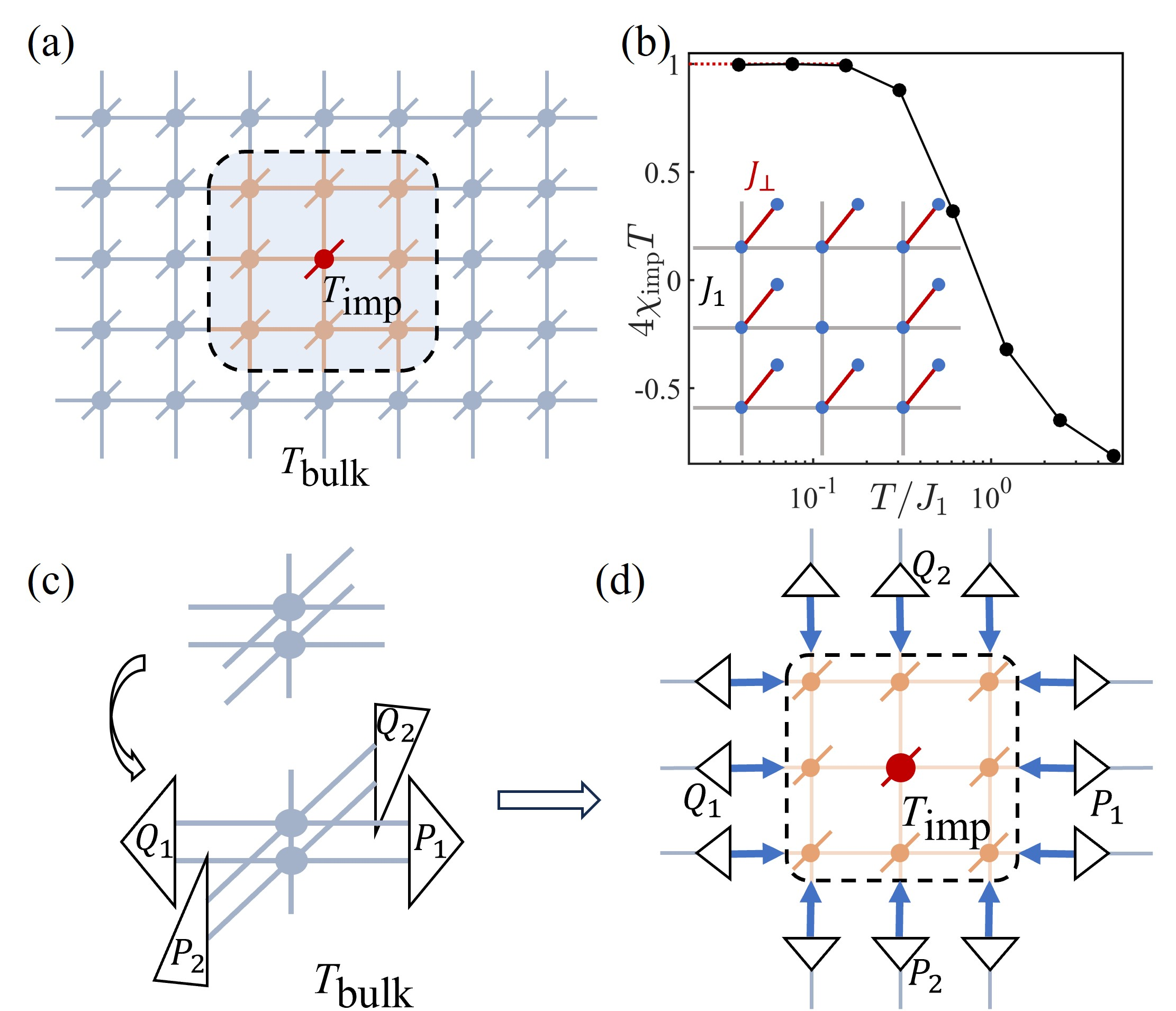}
\caption{\label{fig2} Illustration and benchmark of the IRG method. (a) schematically plots the impurity tensor network ansatz (the shaded region), coupled to the bulk tensors outside this region. (c) Imaginary time-evolution of the density matrix. Truncation of the density matrix is achieved by local transformations and truncations acting upon the virtual bonds of the local tensors, denoted by $P$ and $Q$. (d) The bulk information is transferred into the impurity tensor via $P$ and $Q$ through the impurity-bulk boundary (dashed square). (b) Benchmark on the two-layer spin model with a defect (shown by the inset). $J_1=1$ and $J_{\perp}=2$.}
\end{figure}

\textit{\color{blue}{Impurity renormalization group approach.--}} 
%The first step is to introduce an appropriate tensor-network ansats describing both the interacting baths and the impurity.  
As shown by Fig.~\ref{fig2}(a), we consider a defect (marked by red) locally interacting with the bath defined on a lattice. The defect could either be an external impurity or the removal of a site, located at $\mathbf{r}=0$. Since the translation symmetry of the underlying lattice is broken, the conventional tensor-network ansatz describing many-body wavefunctions with spatial periodicity is no longer applicable. Besides, although matrix product states \cite{Oslund1995, Cirac2007, Cirac2021, GJL} may be used to describe states without translation symmetry \cite{Pollmann}, their accuracy is not guaranteed for 2D systems in the thermodynamic limit \cite{White2012}. 

To overcome these difficulties, we propose an impurity tensor network ansatz, as schematically plotted by Fig.~\ref{fig2}(a). For local (or short-range) impurity-bath interactions, it is natural to separate the whole system into two parts, i.e., the impurity ($\Omega_{\mathrm{imp}}$) and the bulk region ($\Omega_{\mathrm{bulk}}$), separated by the dashed square in  Fig.~\ref{fig2}(a). The former denotes the local area that includes both the impurity itself and the nearby bath degrees of freedom, while the latter is relatively far away from $\mathbf{r}=0$.  For the region away from the defect, the wave function is barely affected. Thus, within $\Omega_{\mathrm{bulk}}$, the translational invariance is effectively maintained, and the wave function can be described by local tensors conventionally defined in a unit cell, i.e.,  $T_{\mathrm{bulk}}$. In sharp contrast, the region  $\Omega_{\mathrm{imp}}$ is strongly influenced by the defect, thus its local tensor should take a completely different form compared to $T_{\mathrm{bulk}}$. Hence, to describe the local wavefunction around the defect, we introduce a cluster of tensors, denoted by $T_{\mathrm{imp}}$, within the region $\Omega_{\mathrm{imp}}$ of the size $\Lambda_{\mathrm{imp}}$, as highlighted by the dashed square in Fig.~\ref{fig2}(a). 

The tensors $T_{\mathrm{imp}}$ and  $T_{\mathrm{bulk}}$ are connected at their boundary via the virtual bonds, forming a whole network describing systems in the thermodynamic limit. Note that in addition to the bond dimensions respectively associated with $T_{\mathrm{imp}}$ and  $T_{\mathrm{bulk}}$, i.e., $\chi_{\mathrm{imp}}$ and $\chi_{\mathrm{bulk}}$, the size of the impurity region, $\Lambda_{\mathrm{imp}}$, provides another numerical parameter. Although the larger $\Lambda_{\mathrm{imp}}$ captures the impurity effects more accurately, the numerical cost also increases, which requires a careful balance, as analyzed in the Supplemental Material \cite{sup}.

Based on the above impurity tensor-network ansatz, we propose a numerical algorithm to obtain the thermodynamic properties induced by the impurity. We first initialize the high temperature density matrix, $\rho_0=e^{-\tau_0H}$, to the tensor form, both for the bulk and impurity region, where $\tau_0=1/T$ is the inverse of temperature. Then, we obtain the density matrix at lower temperatures via imaginary-time evolution. To improve efficiency without losing accuracy, the exponential and linear tensor-network renormalization group (XTRG and LTRG) scheme \cite{WeiLi2011, WeiLi2018} are adopted for the relatively high and low temperature regimes, respectively. 

Due to the presence of two different local tensors, $T_{\mathrm{bulk}}$ and $T_{\mathrm{imp}}$, we introduce the generalized XTRG and LTRG algorithm, i.e., the IRG. Specifically, we first obtain the environment information for $T_{\mathrm{bulk}}$ via the iterative procedure used for tensor networks describing translationally invariant states \cite{Verstraete2008, Gu2008, srg, HHZhao2010, Orus2019finiteT}. Then, $T_{\mathrm{bulk}}$ can be transformed into the quasi-canonical form by inserting a pair of $\chi\times\chi$ unitary matrices, $P$ and $Q$, satisfying $PQ=1$ \cite{WeiLi2012, SM1}, as indicated by Fig.~\ref{fig2}(c). Truncation can then be implemented in the canonical form by cutting off the singular spectrum, $v$, such that the first $\chi^{\star}$ largest values are kept. Then, the truncated matrices $P$, $Q$ and the spectrum $v$ naturally extend their influence to the impurity region. The shared bonds between $\Omega_{\mathrm{bulk}}$ and $\Omega_{\mathrm{imp}}$ inherit either $P$ or $Q$ from the bulk network, inducing corresponding transformations on the bulk-impurity boundary (Fig.~\ref{fig2}(d)). Meanwhile, the vector $v$ transfers the environment information from the bulk to the impurity, establishing the bulk-impurity coupling. In this way, both $T_{\mathrm{bulk}}$ and $T_{\mathrm{imp}}$ are updated, which are used as inputs for the next iteration step. Repeating the above procedures generates the density matrix at lower temperatures and thus the thermodynamic properties.

As a benchmark, we study a two-layer spin-1/2 model with a site vacancy on the top layer, as indicated by the inset in Fig.~\ref{fig2}(b). The Hamiltonian reads as, $H=J_1\sum_{\langle i,j\rangle}\mathbf{S}_{1,i}\cdot\mathbf{S}_{1,j}+J_{\perp}\sum_i\mathbf{S}_{1,i}\cdot\mathbf{S}_{2,i}$, with $J_1,J_{\perp}>0$. The  bottom layer spins are interacting via the Heisenberg term $J_1$, which is absent for the upper layer. The spins of two layers are coupled via the exchange interaction $J_{\perp}$. For large $J_{\perp}$, the spins from the two layers form a singlet state for each site \cite{Wang2006}. Thus, the removal of a site at $\mathbf{r}=0$ locally breaks the corresponding singlet, setting free an effective local moment \cite{Coleman2021}. Behaving as a free spin, the impurity then displays a Curie-like response behavior at low-temperatures. This could be verified by calculating the impurity magnetic susceptibility $\chi_{\mathrm{imp}}$, a key quantity in impurity problems \cite{Fang2015}, defined as $\chi_{\mathrm{imp}}=\chi_{\mathrm{tot}}-\chi$ where $\chi_{\mathrm{tot}}$ and $\chi$ are the magnetic susceptibility without and with the vacancy. As shown in Fig.~\ref{fig2}(b), $T\chi_{\mathrm{imp}}$ saturates to the constant $1/4$ at low-temperatures. This is consistent with the free spin picture, which exhibits $\chi_{\mathrm{imp}}=C/T$ with $C=S(S+1)/3$. The results in Fig.~\ref{fig2}(b) are in agreement with Ref.\cite{Hoglund2007}, justifying the validity of our proposed method. More benchmarks on different models are included in Sec. \uppercase\expandafter{\romannumeral2} of Supplemental Materials.   

\begin{figure}
\includegraphics[width=\linewidth]{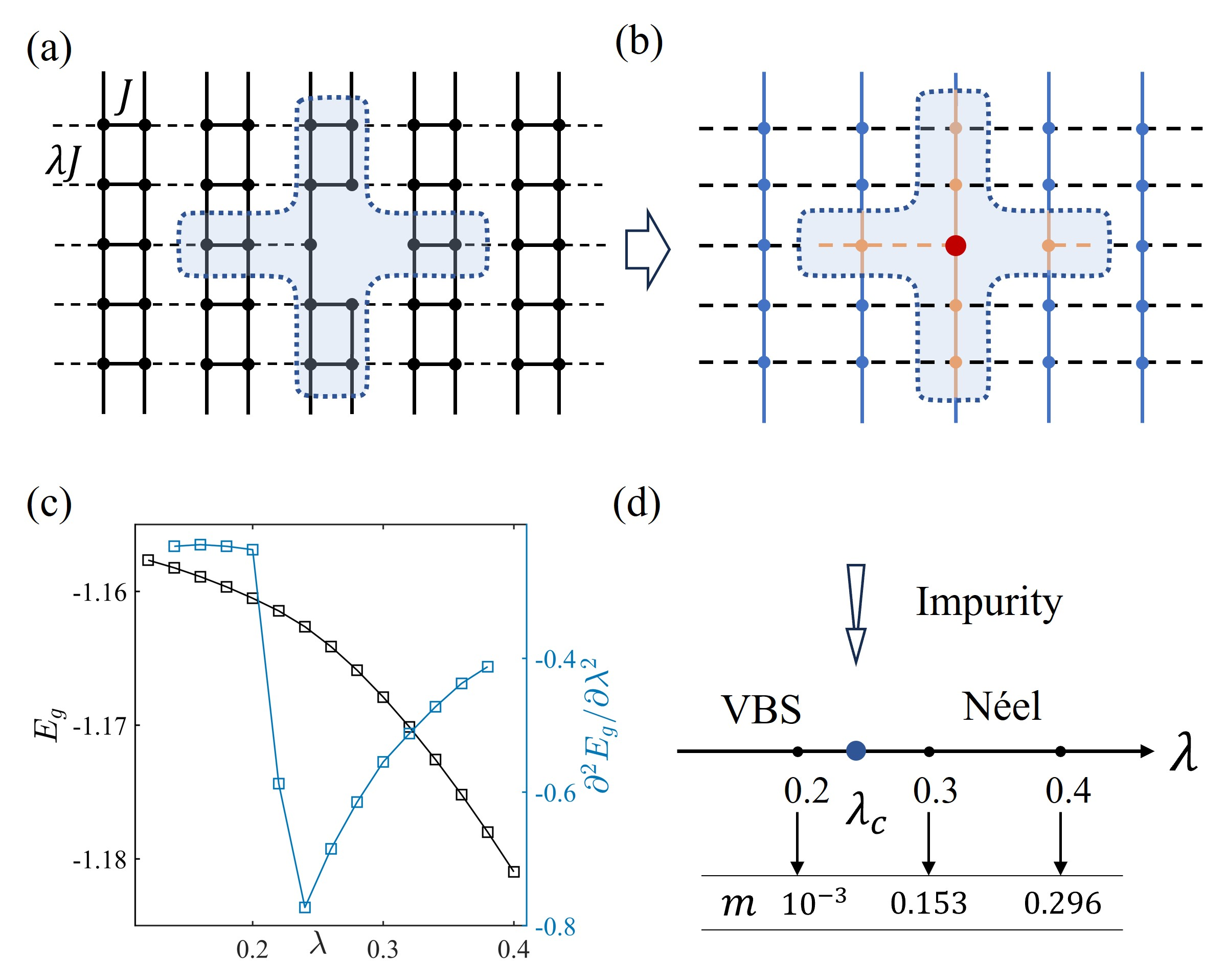}
\caption{\label{fig3} (a) The coupled Heisenberg ladder model with a defect. The shaded region surrounding the defect defines the impurity region. (b) shows the tensor-network representation employed in the calculation, where the spins on each rung are grouped into a single cluster. The shaded region shows the impurity tensor-network ansatz. (c) The ground state energy and its second derivative with respect to $\lambda$ corresponding to the defect-free model. Details are provided in Sec. \uppercase\expandafter{\romannumeral5} of the Supplemental Material. (d) The zero-temperature phase diagram indicated by (c). The magnetization $m$ changes from vanishing to nonzero values passing through the critical point $\lambda_c$. Adding a defect could further probe the transition and its effects at finite-temperatures.}
\end{figure}

\textit{\color{blue}{Impurity probe of quantum critical regime.--}} We now utilize the IRG approach to study the QCR of strongly interacting spin models. We consider a coupled spin-1/2 Heisenberg ladder model on a 2D square lattice with a defect \cite{sachdev1999}, as shown by Fig.~\ref{fig3}(a). The clean system without defects is described by,
\begin{equation}\label{eqmodel}
    H=J\sum_{i,j\in A}\mathbf{S}_i\cdot\mathbf{S}_j+\lambda J\sum_{i,j\in B}\mathbf{S}_{i}\cdot\mathbf{S}_j,
\end{equation}
where $A$ and $B$ denote the intra- and inter-ladder bonds, as marked by the solid and dashed lines in Fig.~\ref{fig3}(a). With increasing $\lambda$ from 0 to 1, the model gradually interpolates between uncoupled ladders and a 2D Heisenberg magnet. In this process, a VBS to an AFM quantum phase transition is expected \cite{sachdev1999, Katoh, Imada}. 

We first investigate the ground state properties at zero temperature for the defect-free case. Performing the imaginary time evolution based on a designed projected entangled pair states (PEPS) \cite{sup}, we calculate the ground state, the corresponding energetics and the magnetization with varying $\lambda$. As shown in Fig.~\ref{fig3}(c), the second derivative of the ground state energy exhibits a singularity at a critical $\lambda_c\sim0.23$, in qualitative agreement with previous studies \cite{Katoh, Imada}. Meanwhile, the magnetization jumps from zero to a nonzero value, clearly indicating the VBS-AFM transition at $\lambda_c$ (Fig.~\ref{fig3}(d)). 

At finite temperatures, the quantum critical point is extended to a critical regime, resulting in crossovers between different regimes, as indicated by Fig.~\ref{fig1}(d). Is it possible to probe the crossover behavior via an impurity, and whether there are any exotic impurity-induced phenomena in the QCR? To address these questions, we remove a single site out of Eq.\eqref{eqmodel} and adopt the tensor-network ansatz shown by Fig.~\ref{fig3}(b).

\begin{figure}
\includegraphics[width=\linewidth]{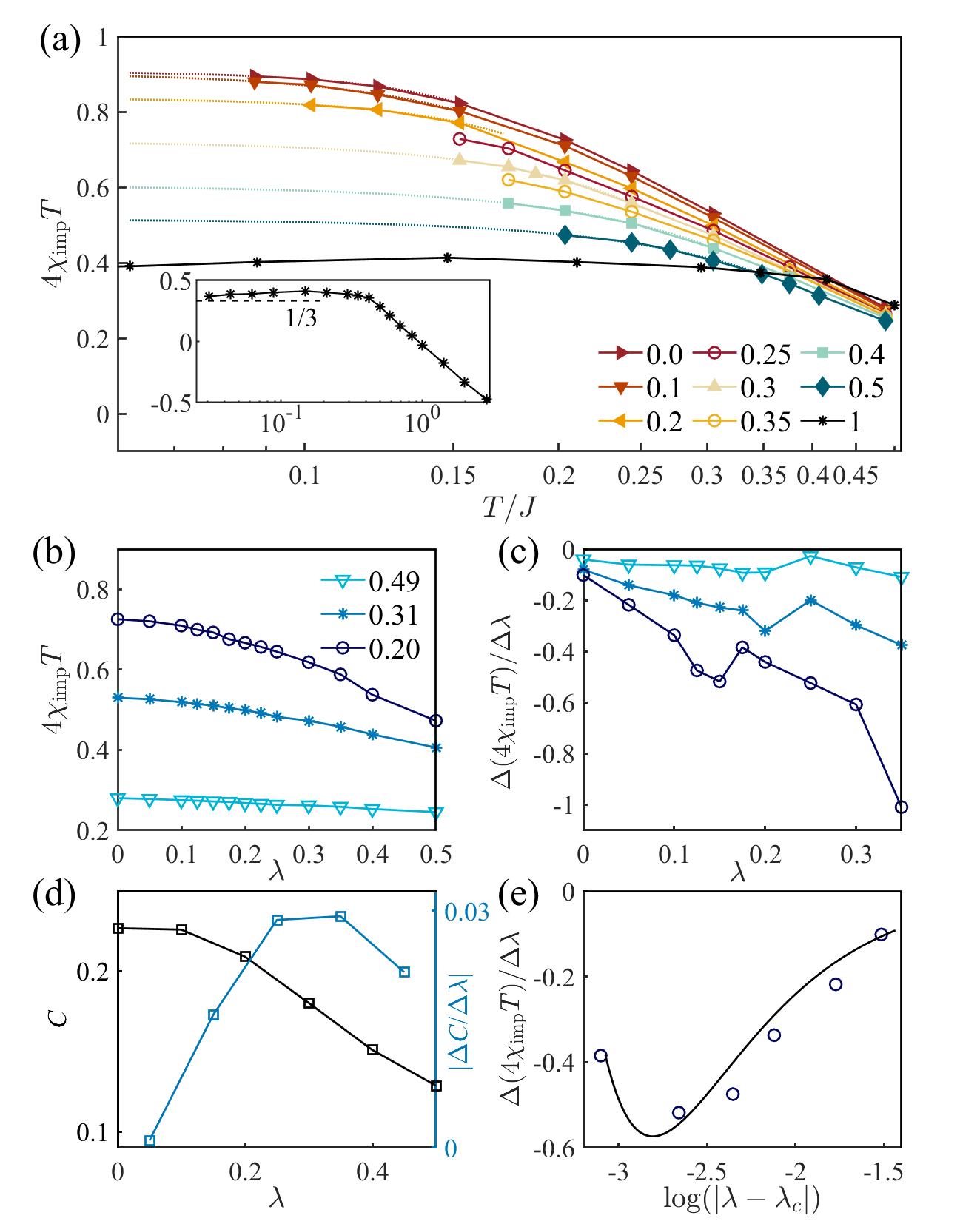}
\caption{\label{fig4} The impurity magnetic susceptibility $\chi_{\text{imp}}$ obtained by the IRG method. (a) shows $4\chi_{\text{imp}}T$ as a function of $T$ for different values of $\lambda$. The inset shows the result for $\lambda = 1$ approaching towards $1/3$ with lowering $T$. The numerical details regarding the bond dimensions are included in Sec. \uppercase\expandafter{\romannumeral4} of Supplemental Materials. (b) shows $4\chi_{\text{imp}}T$ as a function of $\lambda$ for different temperatures $T$. (c) shows the derivative of $4\chi_{\text{imp}}T$ with respect to $\lambda$ for different $T$. (d) plots the low-temperature Curie coefficient $C(\lambda)$ and its derivative with respect to $\lambda$. $C$ decreases down to $1/12$ for $\lambda=1$, as indicated by (a). (e) The data points display the derivative of $4\chi_{\text{imp}}T$ versus $\log(|\lambda-\lambda_c|)$ on the VBS side at $T=0.2$, which is in well consistence with the field-theoretical calculations as shown by the black curve.}
\end{figure}

We then calculate the impurity susceptibility $\chi_{\mathrm{imp}}$ as defined above. Fig.~\ref{fig4}(a) shows $T\chi_{\mathrm{imp}}$ versus temperature $T$, with increasing $\lambda$ encompassing the QCR. In general, we find that $T\chi_{\mathrm{imp}}$ increases with lowering $T$, saturating to constant values (denoted by $C$) at low $T$. Thus, at low temperatures, the impurity always behaves as an effective local moment, displaying Curie-like response, $\chi_{\mathrm{imp}}\sim \frac{C(\lambda)}{T}$. Moreover, $C(\lambda)$ gradually decreases with increasing $\lambda$, exhibiting a smooth evolution in consistence with the crossover behavior \cite{note2}.

Based on effective field theory, Sachdev \textit{et al.} have analyzed the $T\rightarrow0$
asymptotic impurity susceptibility \cite{sachdev1999}.  For the VBS, it was argued that an effective $S=1/2$ spin is set free by removing a single site, giving rise to $T\chi_{\mathrm{imp}}=S(S+1)/3$  for  $\lambda<\lambda_c$. Interestingly, with increasing the size of the impurity tensor $\Lambda_{\mathrm{imp}}$, we do observe that $C(\lambda=0)$ approaches the predicted value $1/4$ (for $S=1/2$), as shown in Sec. \uppercase\expandafter{\romannumeral2} of Supplemental Materials. In the AFM phase, the defect generates a total magnetic moment quantized at $S$, which behaves as a classical spin because of the locking of the moment orientation to the local N\'{e}el order. Thus, the impurity exhibits the longitudinal susceptibility dominated by $T\chi_{\mathrm{imp}}\approx S^2/3$  for  $\lambda>\lambda_c$. This behavior has been verified by quantum Monte Carlo simulations at the Heisenberg point $\lambda=1$ \cite{sandvik2003}, which found a leading value of $C(\lambda=1)=1/12$ for $S=1/2$ (accompanied by logarithmic corrections), as shown by the inset in Fig. \ref{fig4}(a).  However, numerical studies of the evolution process towards the AFM have been missing to date. Moreover, at the critical point $\lambda=\lambda_c$, the renormalization group (RG) calculation has yielded $\chi_{\mathrm{imp}}=\tilde{S}(\tilde{S}+1)/3T$, where $\tilde{S}\neq S$  is an emergent fractional spin \cite{sachdev1999, sandvik2003}. Although the emerging spin fractional induced by defects has drawn significant interest for decades,  its numerical evidence in the coupled ladder model and how it evolves within the crossover region are yet to be addressed. These will be answered by our impurity renormalization group approach in the following. 

We extract the low-temperature Curie coefficient $C(\lambda)$ as shown in Fig.~\ref{fig4}(d). Clearly, it  gradually evolves from values close to $1/4$ towards $1/12$  with increasing $\lambda$.  As illustrated above, both $C=1/4$ and $1/12$ reflect the quantized moment around the defect, i.e.,  $S=1/2$. However, in the intermediate $\lambda$ region, generic values between $1/4$ and $1/12$ clearly indicate the emergence of a fractionalized local moment. In addition, at the critical point $\lambda\sim0.23$, we obtain $\tilde{S}\sim 0.19$. which is  comparable to the value $0.16$ obtained by RG calculations based on $\epsilon$-expansion \cite{sachdev2000}. Therefore, our impurity tensor-network approach not only verifies the field-theoretical asymptotic results but also clearly observes the defect-induced fractionalized local moment. Notably, such a fractionalization arises from the interplay between the bath correlation and the defect.

We now examine the crossover behaviors towards the quantum critical regime with tuning $\lambda$ and fixing $T$. As shown in Fig.~\ref{fig4}(b), $T\chi_{\mathrm{imp}}$ decreases  with $\lambda$ for various temperatures. Moreover, taking the derivative of $T\chi_{\mathrm{imp}}$ with respect to $\lambda$, we find emergence of a non-monotonous evolution behavior with raising $\lambda$ (Fig.~\ref{fig4}(c)), resulting in cusps which become more manifested for lower temperatures.

To fully understand the above temperature dependence of the impurity susceptibility, we adopt the field-theoretical approach in Ref.~\cite{sachdev2000}. The spin fluctuations underlying the VBS-AFM transition of the coupled ladder model is described by $S_{\mathrm{bulk}}=\int^{\beta}_0\int d^2x[(\partial_{\mu}\phi_{\alpha})^2+\tilde{\lambda}\phi^2_{\alpha}+u(\phi_{\alpha}\phi_{\alpha})^2]$,  where $\phi_{\alpha}$ is the vector AFM order parameter, and the index $\mu=(\tau,x_1,x_2)$ involves both the imaginary time and spatial coordinate. The modification introduced by defect is described by $S_{\mathrm{imp}}$, which includes an uncompensated Berry phase term as well as the coupling at the impurity site, $-\gamma S\int^{\beta}_0d\tau\phi_{\alpha}(\tau)n_{\alpha(\tau)}$, where the vector $n_{\alpha}(t)$ denotes the orientation of the defect induced local moment, and $\gamma$ denotes its coupling with the bath order parameter. In the RG sense, the beta functions of $u$ and $\gamma$ can be derived, and the impurity susceptibility is obtained by calculating the renormalized correlation functions order by order. Focusing on the VBS side, i.e. $\lambda\lesssim\lambda_c$, it is found that \cite{sachdev2000}, 
\begin{equation}\label{eqscaling}
    T\chi_{\mathrm{imp}}=\frac{S(S+1)}{3}[1+\Phi^{(1)}_{\mathrm{imp}}+\Phi^{(2)}_{\mathrm{imp}}+...],
\end{equation}
where $\Phi^{(1)}_{\mathrm{imp}}\sim\gamma^2T\int\frac{d^2k}{(2\pi)^2}\frac{1}{\varepsilon^4_k}$, $\epsilon_k=\sqrt{k^2+m^2}$ and $m\sim\epsilon^{1/2}T$ for large $T$. This term belongs to the $O(\epsilon^{1/2})$ order in the $\epsilon$-expansion.   The second term in Eq.~\eqref{eqscaling} is of the order $O(\epsilon)$, which is derived as $\Phi^{(2)}_{\mathrm{imp}}\sim-\gamma^2u[\int\frac{d^2k}{(2\pi)^2}\frac{1}{\varepsilon^4_k}]^2$. Note that the negative sign in front of $\Phi^{(2)}_{\mathrm{imp}}$ indicates a competition with $\Phi^{(1)}_{\mathrm{imp}}$.
\begin{figure}
\includegraphics[width=\linewidth]{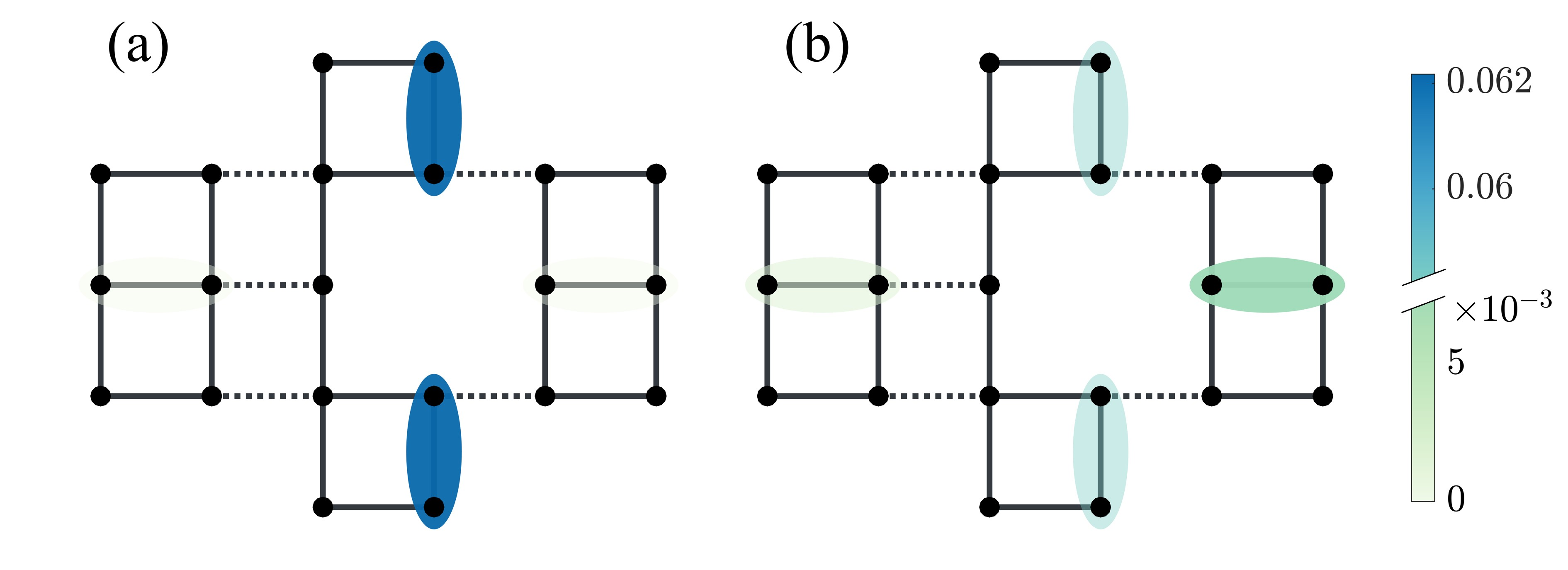}
\caption{\label{fig5} The impurity-induced spin-spin correlation function around the defect obtained at a  fixed temperature ($T = 0.31$). With increasing $\lambda$, the correlation texture exhibits a columnar to pinwheel-like evolution. $\lambda = 0$ and $0.4$ for (a) and (b), respectively.}
\end{figure}

Although $T\chi_{\mathrm{imp}}$ was evaluated at the critical point $\lambda_c$ in Ref.~\cite{sachdev2000}, the $O(\epsilon^{1/2})$ and $O(\epsilon)$ order contributions way from the critical point and how they compete with each other are unclear. For analytic convenience, we consider the VBS side by reducing $\lambda$ from $\lambda_c$. The VBS spin gap, i.e., $\Delta$, increases as $\Delta\sim|\lambda-\lambda_c|^{\nu}$, with $\nu$ being a known exponent. Meanwhile, the renormalized mass $m$ interpolates to $m\sim\Delta$ on the VBS side for $T\ll\Delta$ via a crossover function \cite{sachdev1997}. Thus, the ``distance" to the critical point, $\lambda-\lambda_c$, is parametrized by $ m$, as $\log m\sim\nu\log|\lambda-\lambda_c|$. We thereby numerically evaluate $T\chi_{\mathrm{imp}}$ from Eq.~\eqref{eqscaling} and plot its derivative as a function of  $\log m$. The results are shown by the black curve in Fig.~\ref{fig4}(e), which well fits the data obtained by our IRG approach. It therefore becomes clear that the non-monotonous  feature  in Fig.~\ref{fig4}(c) is physically originated from the competition between the $O(\epsilon^{1/2})$ and $O(\epsilon)$ contributions, which is non-negligible around the crossover region. Thus, with increasing $\lambda$ while fixing $T$, such a cusp in the derivative of the impurity susceptibility could act as an efficient signal that probes the crossover towards the quantum critical regime. 

Last,  we plot in Fig.~\ref{fig4}(d) the derivative of the low-$T$ Curie coefficient with respect to $\lambda$. Similar to  Fig.~\ref{fig4}(c), the cusp-like feature is found, whose plateau maximum locates around the critical point $\lambda_c$. This implies that the $O(\epsilon^{1/2})$-$O(\epsilon)$ competition persists at low-temperatures, reflecting the essential low-energy fluctuations around the quantum critical point.

\textit{\color{blue}{Evolution of impurity induced correlation texture.--}} The field-theoretical analysis above is unable to capture the detailed, short-range properties induced by the defect, which is remedied by our IRG approach. To demonstrate this, we calculate the spin correlation functions between the nearest sites $i$ and $j$, i.e., $C_{ij}=\langle S_iS_j\rangle$, for the case with and without the defect, respectively. Then, we plot the change of $C_{ij}$ induced by the impurity, i.e., $C_{\mathrm{imp},ij}$. As shown by Fig.~\ref{fig5}, for $\lambda=0$ in the VBS phase, $C_{\mathrm{imp},ij}$ is mainly distributed on the two vertical bonds nearby the defect (Fig.~\ref{fig5}(a)). With increasing $\lambda$ towards the QCR, we find that $C_{\mathrm{imp},ij}$ on the two vertical bonds are gradually weakened. Meanwhile, the correlations on the two horizontal bonds are enhanced, resulting in a pinwheel-like correlation texture for large $\lambda$, as shown by Fig.~\ref{fig5}(b). Thus, we observe  a continuous evolution of the spin-spin correlation regarding its real-space texture, which is beyond previous approaches. Such a columnar to pinwheel-like evolution could be experimentally feasible, e.g., via spin-polarized STM/STS measurements.   

\textit{\color{blue}{Discussion and conclusion.}}--
%Although we demonstrate the IRG method by investigating the QCR, it can be straightforwardly applied to solve a number of long-standing problems, including the vacancies in spin liquids [Cite], the defects in deconfined quantum criticalities [Cite], and the impurities in non-Fermi liquid hosts [Cite]. To completely solve these problems, it is necessary to simulate the low-energy excitations of the correlated bath and meanwhile properly treats their interaction with the defect. This critical challenge is well resolved by our IRG method, opening new possibilities to discover novel impurities states previously inaccessible. In addition, it is also interesting to simulate the zero-temperature impurity ground state using our proposed tensor-network ansatz. Moreover, the accuracy could be further enhanced by using the full update or the variational methods. 
Although we demonstrate the efficiency of our IRG method by investigating the QCR, it can be directly applied to solve several long-standing problems. These include vacancies in quantum spin liquids \cite{Willans2010, Takahashi2023, Yatsuta2024}, defects in deconfined quantum critical points \cite{DQCP, shu2025}, and impurities in non-Fermi liquid systems \cite{RW2019, RW2021, JXZ2001}. A complete solution to these problems requires simultaneously simulating the low-energy excitations of the correlated baths and their interaction with the defect---a key challenge that our IRG method resolves. This opens new avenues for discovering novel impurity states that were previously inaccessible. Furthermore, the zero-temperature impurity ground state can also be simulated using our proposed tensor-network ansatz, and the accuracy of which can be further improved via full-update or variational methods.

%Thus, the IRG points to a new direction avenue is opened, towards a comprehensive framework where the tensor-network is accomadated with defects witout translational invariance.

\begin{acknowledgments}
R. W. acknowledges Tigran Sedrakyan for fruitful discussions. This work was supported by the Innovation Program for Quantum Science and Technology (Grant no. 2021ZD0302800), the National R\&D Program of China (2022YFA1403601), the National Natural Science Foundation of China (No. 12322402, No. 12274206), the Natural Science Foundation of Jiangsu Province (No. BK20233001), and the Xiaomi foundation.
\end{acknowledgments}


\begin{thebibliography}{99}
\bibitem{note3} Here, ``non-interacting" involves the cases where the reservoirs are treated in mean-field level and behave effectively as non-interacting systems in the quasi-particle basis, such as the superconductors.
\end{thebibliography}
\end{document}

\begin{center}
\textbf{\large Supplemental material for}
\end{center}

\title{Probing quantum critical crossover via impurity renromalization group}
\author{Tao Yang}
\affiliation{National Laboratory of Solid State Microstructures and Department of Physics, Nanjing University, Nanjing 210093, China}

\author{Z. Y. Xie}
\email{qingtaoxie@ruc.edu.cn}
\affiliation{School of Physics, Renmin University of China, Beijing 100872, China}
\affiliation{Key Laboratory of Quantum State Construction and Manipulation (Ministry of Education), Renmin University of China, Beijing 100872, China}

\author{Rui Wang}
\email{rwang89@nju.edu.cn}
\affiliation{National Laboratory of Solid State Microstructures and Department of Physics, Nanjing University, Nanjing 210093, China}
\affiliation{Collaborative Innovation Center of Advanced Microstructures, Nanjing University, Nanjing 210093, China}
\affiliation{Jiangsu Physical Science Research Center}
\affiliation{Hefei National Laboratory, Hefei 230088, People's Republic of China }

\author{Baigeng Wang}
\affiliation{National Laboratory of Solid State Microstructures and Department of Physics, Nanjing University, Nanjing 210093, China}
\affiliation{Collaborative Innovation Center of Advanced Microstructures, Nanjing University, Nanjing 210093, China}
\affiliation{Jiangsu Physical Science Research Center}

\def\tred#1{{\textcolor{red}{#1}}}
\captionsetup[figure]{name=Fig.}
\renewcommand{\thefigure}{S\arabic{figure}}
\renewcommand{\thetable}{S\arabic{table}}
\renewcommand{\theequation}{S\arabic{equation}}
% \captionsetup{justification=raggedright, singlelinecheck=false}

\begin{abstract}
We provide supplemental information regarding all technical details on the proofs and derivations of key conclusions presented in the manuscript.  Specific contents include: 1) Numerical algorithm, 2) Benchmark on different models, 3) Dependence on the geometric shape of the impurity ansatz, 4) Convergence of numerics, and 5) Technical details on solving the ground state.
\end{abstract}
\maketitle

\section{Numerical algorithm}\label{Sec:algorithm}
In the main text, we propose the IRG approach, which enables a systematic description of the imaginary-time evolution of impurities embedded in interacting baths in the thermodynamic limit. In this section, we provide a detailed demonstration of this approach.

The method starts from representing the density matrix by the tensor network at high temperatures. When the temperature $T = 1/\beta$ is high, $\beta$ is small, allowing $e^{-\beta H}$ to be accurately approximated up to a Trotter error of order $O(\beta^2)$. To lower the temperature and probe the thermodynamic behavior in the lower temperature regime, the tensor network representing the density matrix is evolved in imaginary time through successive steps. 

For translationally invariant systems in the thermodynamic limit, the system’s properties can be efficiently captured by a minimal set of tensors. However, when an impurity is present in the system, its presence perturbs surrounding tensors during imaginary-time evolution, with the perturbation gradually decaying over space. Such an effect requires a rapidly growing number of tensors for its complete description, rendering the large-scale simulations intractable. To remedy this issue, we propose the IRG approach.

\begin{figure}[htbp]
\centering \hspace{25mm}
\includegraphics[width=0.5\textwidth]{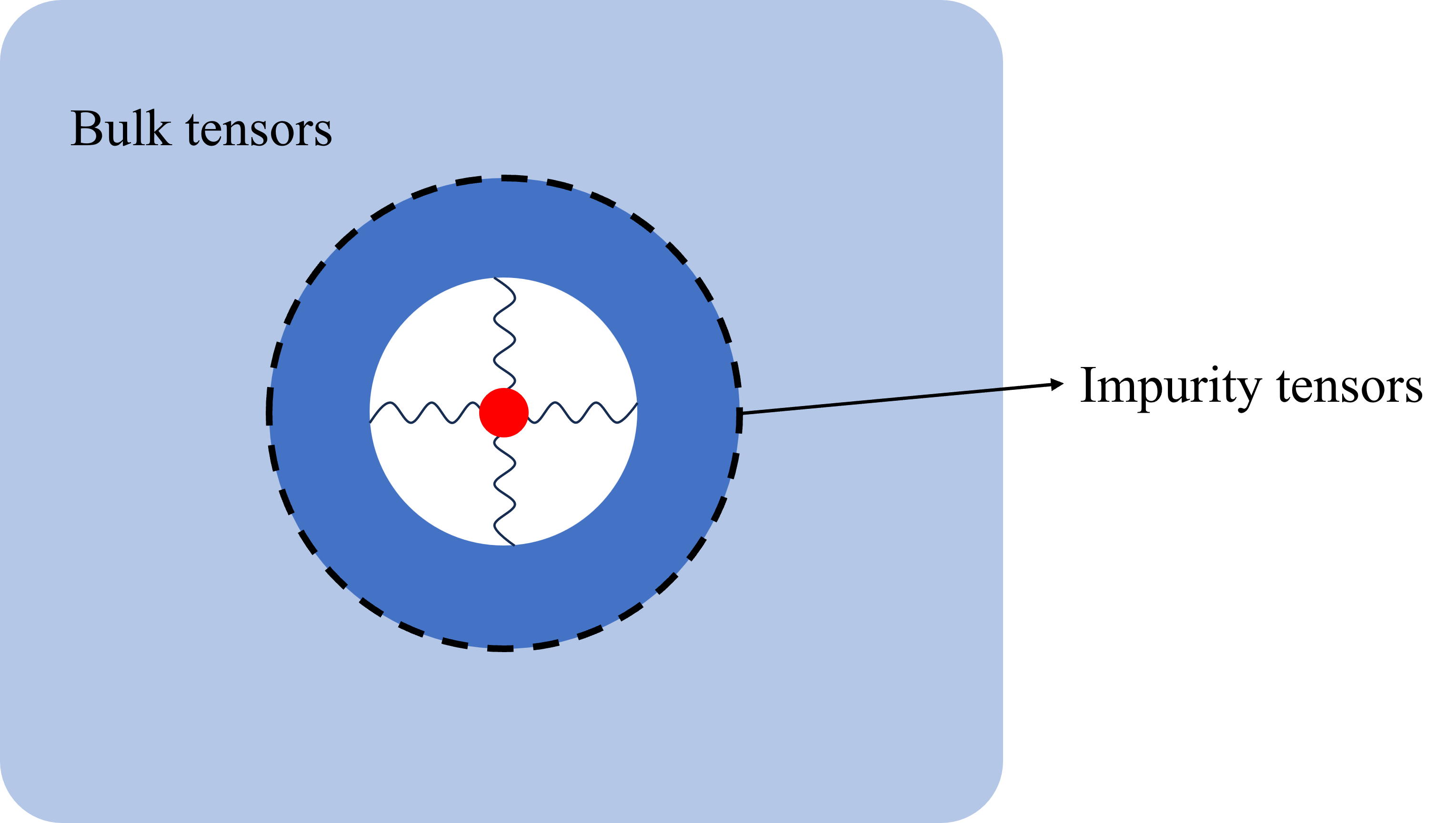}
\caption{\label{fig:imp_tensor}Schematic plot of the tensor-network structure underlying the IRG method.}
\end{figure}

The IRG approach introduces an $\textit{impurity tensor-network ansatz}$, i.e, a cluster of tensors encompassing the impurities are designated as impurity tensors, surrounded by bulk tensors, as shown by Fig.~\ref{fig:imp_tensor}. This establishes a spatial boundary, separating the impurity and bulk region. During imaginary-time evolution, the influence of impurity is confined to this impurity tensor regime, as indicated by Fig.~\ref{fig:imp_tensor}. Below, we demonstrate how the imaginary evolution is implemented for the bulk and impurity tensors.

\begin{figure}[htbp]
\centering
\includegraphics[width=0.9\textwidth]{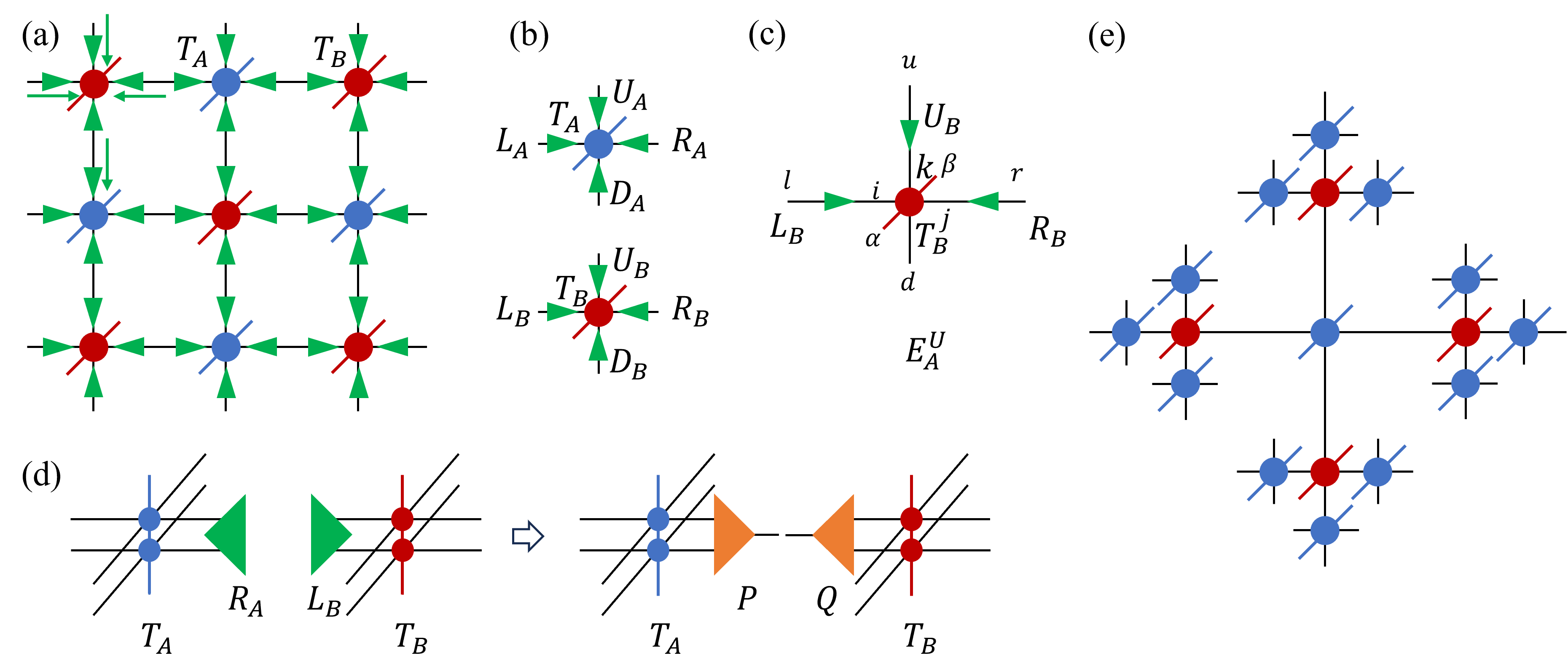}
\caption{\label{fig:Bethe}Bethe canonicalization and truncation of bulk tensors during the XTRG evolution.}
\end{figure}
To simulate the thermodynamic properties in the relatively higher temperature region, the exponential thermal tensor-network (XTRG) method is employed \cite{XTRG}. Starting from the tensor-network representing the density matrix, two identical density matrices are contracted as $\rho_{k+1} = \rho_k \cdot \rho_k^\dagger$ at each step. Correspondingly, the temperature is lowered from $T_k = 1/(2^{k}\tau_0)$ to $T_{k+1} = 1/(2^{k+1}\tau_0)$.  Such a direct contraction expands the virtual bond dimension from $\chi$ to $\chi^2$, necessitating further truncation of the local tensors. For bulk tensors where loops prevent standard canonicalization, we implement the Bethe lattice ansatz for the tensor-network \cite{WeiLi2012}, as shown in Fig.~\ref{fig:Bethe}(e).

The first step of the scheme is to obtain the environment information for each tensor in matrix form. Given that the bulk tensors exhibit translational invariance in the thermodynamic limit, the environment matrix for each tensor index can be obtained through an iterative procedure, where the environment information propagates through the tensor network. Since the environment matrices are dependent on one another, the propagation process leads to a set of self-consistent equations. As illustrated in Fig.~\ref{fig:Bethe}, for the two-sublattice cases, the environment matrices (green triangles), sensed by each virtual bond, participate in these equations. For instance, the upper environment matrix of tensor $T_A$, represented as $U_A$, is related to $T_B$ and other environment matrices, as shown in Fig.~\ref{fig:Bethe}(a)-(c). Specifically,
\begin{align}
    E_{A,lru\alpha\beta d}^U &= \sum_{i,j,k} T_{B,ijkd\alpha\beta} L_{B,li} R_{B,rj} U_{B,uk}, \label{eq1} \\
    E_{A,lru\alpha\beta d}^U &= \sum_m Q_{lru\alpha\beta m} U_{A,md}. \label{eq2}
\end{align}
Here, the Eq.~\eqref{eq1} demonstrates how $E_{A}^U$ absorbs the environmental contribution ($L_B$, $R_B$, $U_B$) mediated by $T_B$, while the Eq.~\eqref{eq2} reflects the QR-based update rule for $U_A$. 
These self-consistent equations enable iterative updates, driving all environment matrices toward convergence. Remarkably, this iteration process avoids the involvement of lattice loops, so it is equivalent to the environment propagation on Bethe lattices. 

After obtaining the environment matrices, the tensor network can be canonicalized in a manner analogous to that in matrix product states \cite{Orus2008}. The canonicalization is implemented by inserting a pair of $\chi \times \chi$ unitary matrices $P* Q$ at each bond between two local tensors, where the adjacent tensors are transformed via contraction with $P$ and $Q$ respectively \cite{WeiLi2012}. For example, for the bond associated with $R_A$ and $L_B$, we perform the singular value decomposition (SVD) to obtain $L_B*R_A=U_c\Lambda_cV_c^T$. From this decomposition, we derive,
\begin{equation}
    P = R_AV_c\lambda_c^{-1/2}, Q = \Lambda_c^{-1/2}U_c^TL_B,
\end{equation}
where $P$ and $Q$ serve as the transformation matrices on the right bond of $T_A$ and the left bond of $T_B$, respectively, as illustrated by Fig.~\ref{fig:Bethe}(d). This process yields the singular spectrum $v = \Lambda_c$ across each bond while establishing the canonical form. The bond dimension reduction from $\chi$ to $\chi^*$ is achieved through the truncation of these transformation matrices, namely the matrix $P$ is reduced from $\chi \times \chi$ to $\chi \times \chi^*$ while $Q$ is truncated to $\chi^* \times \chi$. Therefore, the bulk tensors are effectively compressed.

\begin{figure}[htbp]
\centering
\includegraphics[width=0.5\textwidth]{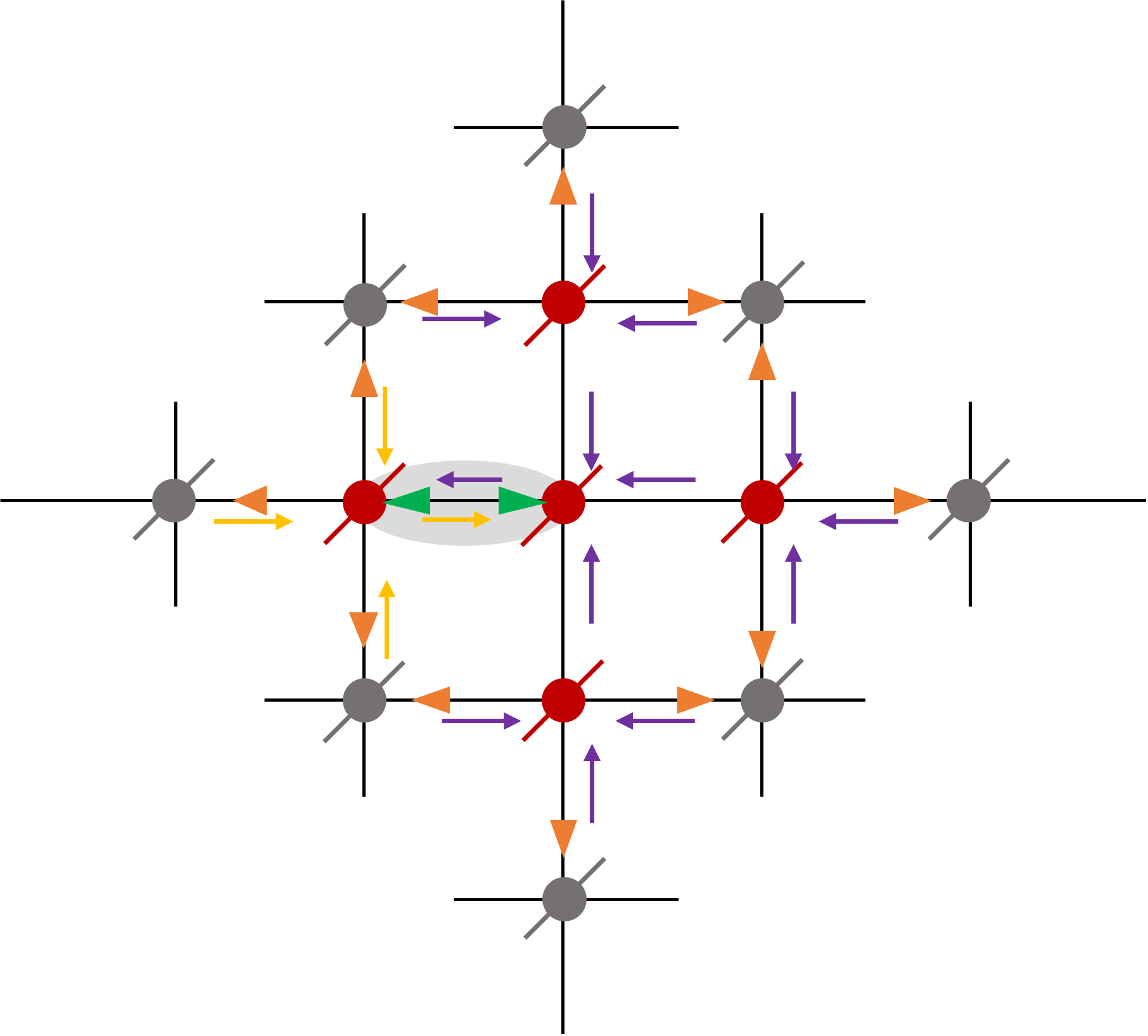}
\caption{\label{fig:imp_cano}The environment information transfer pathway for a specific bond marked by the gray oval.}
\end{figure}

Then, we need to truncate the impurity tensors surrounded by the bulk tensors. The shared bonds between the impurity and bulk tensors inherit either $P$ or $Q$ from the bulk network, inducing the corresponding transformations on the boundary of the impurity tensors. Meanwhile, the environmental information is propagated into the impurity region through the boundary, which determines the bond environment matrices within the impurity tensors. This naturally implements the canonicalization and truncation of the impurity tensors. As an example, we show in Fig.~\ref{fig:imp_cano} the transfer path of the environmental information for a specific bond.

At low temperatures, to collect detailed data at more temperature values, the linearized tensor renormalization group (LTRG) approach is utilized \cite{LTRG}. Unlike the XTRG, where $\beta$ grows exponentially during the evolution process, LTRG implements the imaginary-time evolution of the density matrix by successively projecting new layers of $\rho(\tau)$ onto it. In this process, the inverse temperature $\beta$ increases linearly. LTRG employs the same transformation and truncation strategy as demonstrated above, despite some slight differences in terms of their evolution.

\section{Benchmark on different models}\label{Sec:benchmark}
\begin{figure}[htbp]
\centering
\includegraphics[width=0.8\textwidth]{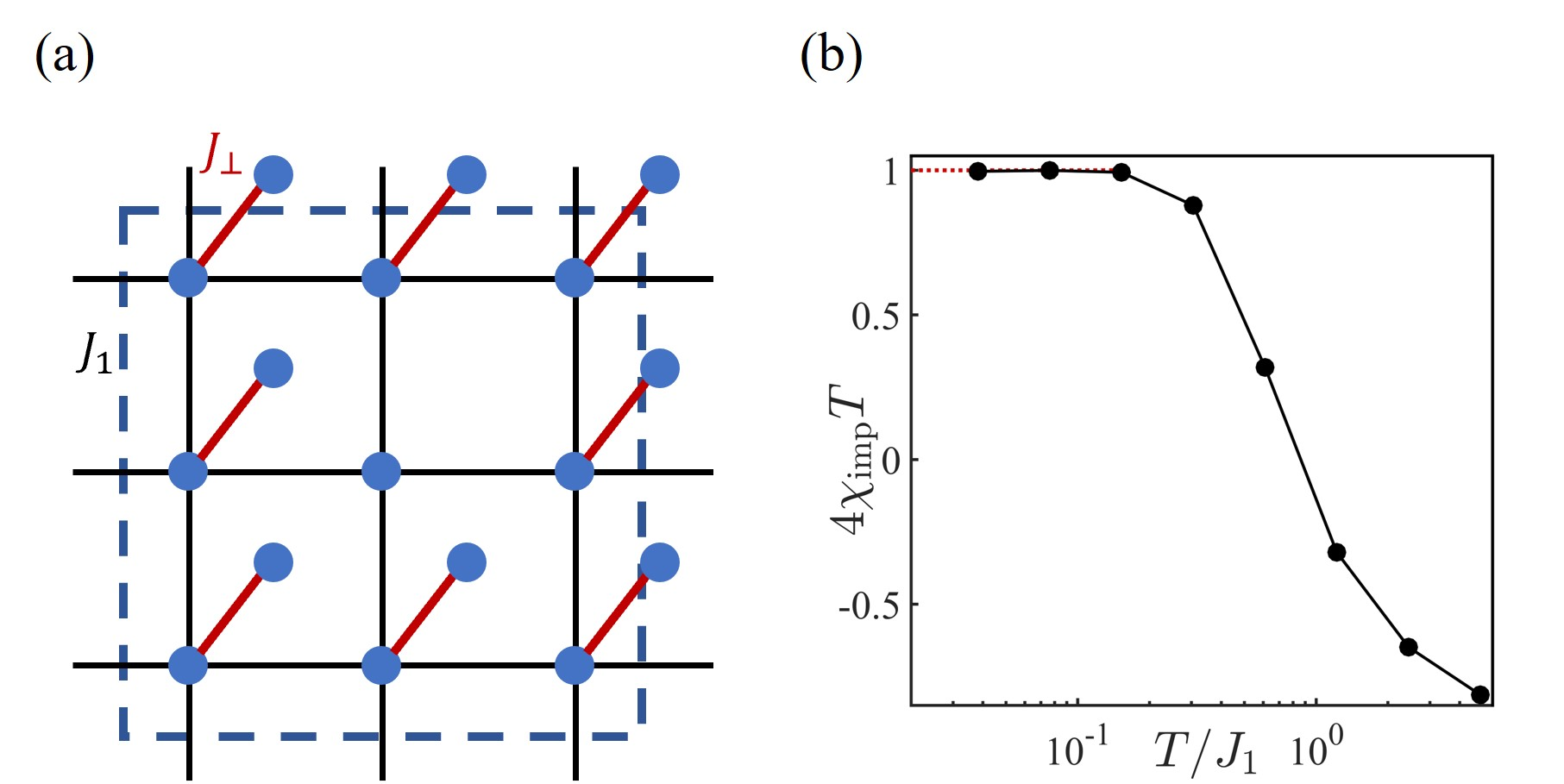}
\caption{\label{fig:incomplete}(a) The Heisenberg model on an incomplete bilayer with a vacancy, where the intra-layer exchange interactions ($J_1$) are represented by the black bonds and the inter-layer interactions ($J_2$) by the red bonds. The impurity tensor-network ansatz is highlighted within the area enclosed by the blue dashed lines, and the bulk tensor component is omitted for simplicity. (b) The impurity magnetic susceptibility versus temperature calculated for the bond dimension 8.}
\end{figure}
In this section, we show benchmarks of the IRG approach on two different impurity models.

The first is a two-dimensional Heisenberg model on an incomplete bilayer lattice containing a vacancy, which is briefly discussed in the main text, as illustrated in Fig.~\ref{fig:incomplete}(a). Here, `incomplete' refers to the lattice configuration where only the bottom layer exhibits intra-layer interactions. The Hamiltonian reads as
\begin{equation}
    H = J_1 \sum_{\langle i, j \rangle} \boldsymbol{S}_{1, i} \cdot \boldsymbol{S}_{1, j} + J_\perp \sum_i \boldsymbol{S}_{1, i} \cdot \boldsymbol{S}_{2, i}.
\end{equation}
The spin-1/2 operator $\boldsymbol{S}_{\alpha, i}$ denotes the spin at site $i$ in layer $\alpha$, where $\langle i, j \rangle$ represents the nearest-neighbor bonds. In this model, the intra-layer coupling strength $J_1$ is present only in the bottom layer $\alpha=1$, while the inter-layer coupling is determined by $J_{\perp}$.

Previous studies~\cite{Wang2006} have demonstrated that the system undergoes a quantum phase transition at $g_c = J_\perp / J_1 = 1.3888(1)$, from N\'{e}el order for $g < g_c$ to the spin-gapped VBS phase for $g > g_c$. In the VBS regime, the intra-layer spins at the same site form strong inter-layer dimers. Upon removing a spin in the top layer $\alpha = 2$, its dimer partner is decoupled, leaving only weak intra-layer correlations. This results in the confinement of an effective local $S = 1/2$ magnetic moment in the vicinity of the vacancy \cite{Sachdev1999}. At low temperatures, the effective net moment contributes a Curie-like behavior to the impurity magnetic susceptibility, i.e.,
\begin{equation}
    \chi_{\text{imp}} = \frac{S(S+1)}{3k_BT}. \label{Curie}
\end{equation}

Using the IRG approach, we investigate the incomplete bilayer model at $g = 2$, and calculate the impurity magnetic susceptibility $\chi_{\text{imp}}$. The impurity tensor-network ansatz adopted is depicted in Fig.~\ref{fig:incomplete}(a), and it is shown in Fig.~\ref{fig:incomplete}(b) that, when $T/J_1\lesssim0.15$, $\chi_{\text{imp}} \approx 1/(4T)$, in consistence with previous results.

\begin{figure}[htbp]
\centering
\includegraphics[width=0.8\textwidth]{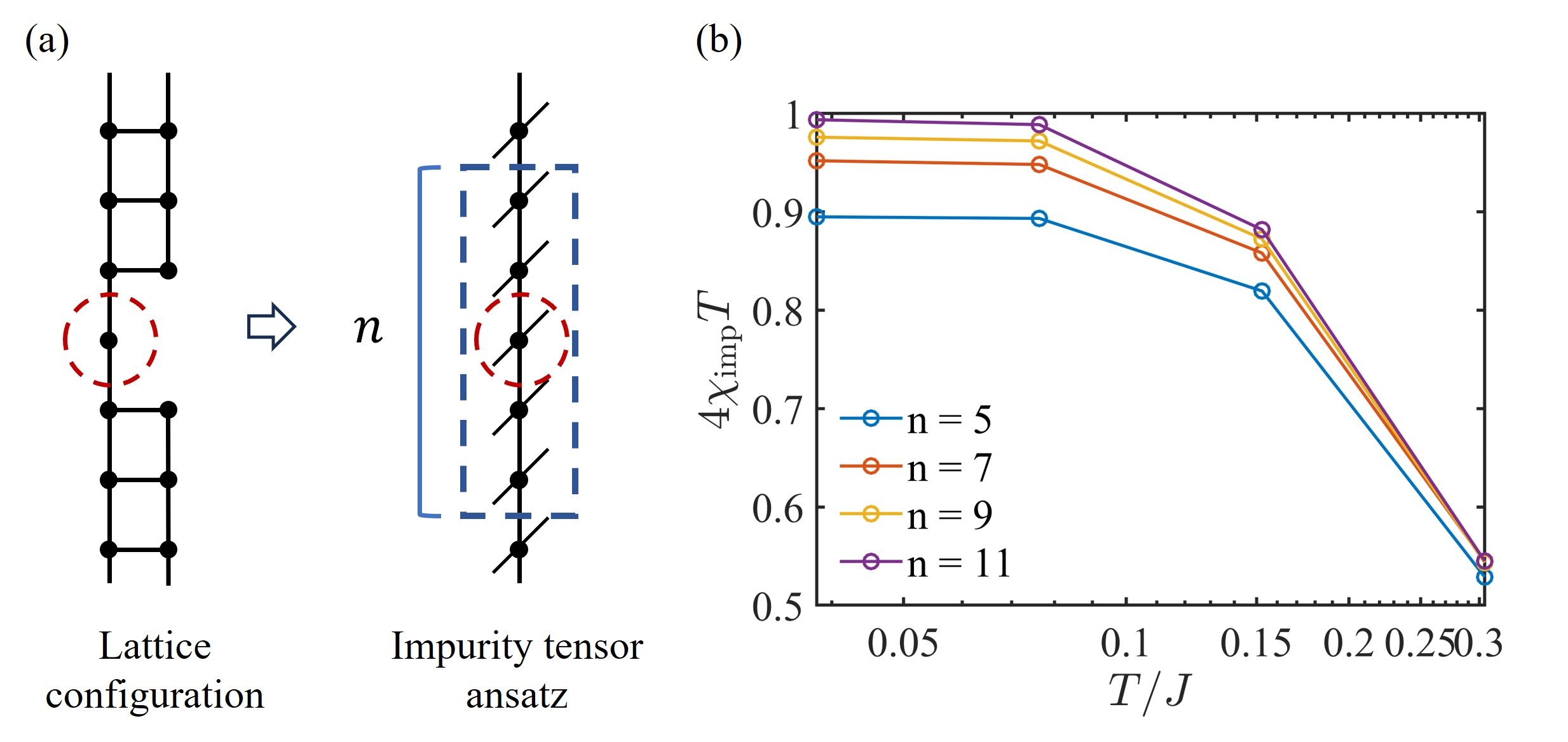}
\caption{\raggedright\label{fig:ladder}(a)Left: the Heisenberg ladder model with a vacancy. Right: the impurity tensor-network ansatz with the parameter $n=5$ (denoting the size of the impurity ansatz in the main text, $\Lambda_{\mathrm{imp}}$). (b) The impurity magnetic susceptibility as a function of temperature for different ansatz sizes, where the bond dimension of 30 is used for the calculation.}
\end{figure}

The second model is the one-dimensional Heisenberg ladder with a vacancy. The bulk Hamiltonian without impurity is described by,
\begin{equation}
    H = J\sum_\alpha \sum_{\langle i,j \rangle}\boldsymbol{S}_{\alpha, i}\cdot \boldsymbol{S}_{\alpha, j}+J\sum_i \boldsymbol{S}_{A, i} \cdot \boldsymbol{S}_{B, i},
\end{equation}
where $\boldsymbol{S}_{\alpha, i}$ is a spin-1/2 at the rung position $i$, with the leg index $\alpha$. This ladder model corresponds to the coupled Heisenberg ladder model discussed in the main text in the specific limit $\lambda = 0$. The ground state of this model exhibits the VBS phase. When a vacancy is introduced into this ladder, the same behavior as in Eq.~\ref{Curie} is expected.

Unlike the incomplete bilayer model at high $g$ values, where interlayer dimer correlations dominate, this model exhibits stronger spatial correlations with a longer correlation length. Thus, the result of $\chi_{\text{imp}}$ is quantitatively affected by the size of the impurity ansatz, i.e., the length $n$ displayed in Fig.~\ref{fig:ladder}(a). When $n$ is increased, $T\chi_{\mathrm{imp}}$ approaches the predicted value $1/4$ more and more closely. The results on the above two models justify the applicability of the IRG method to quantum spin models with defects. 

\begin{figure}[t]
\centering
\includegraphics[width=0.8\textwidth]{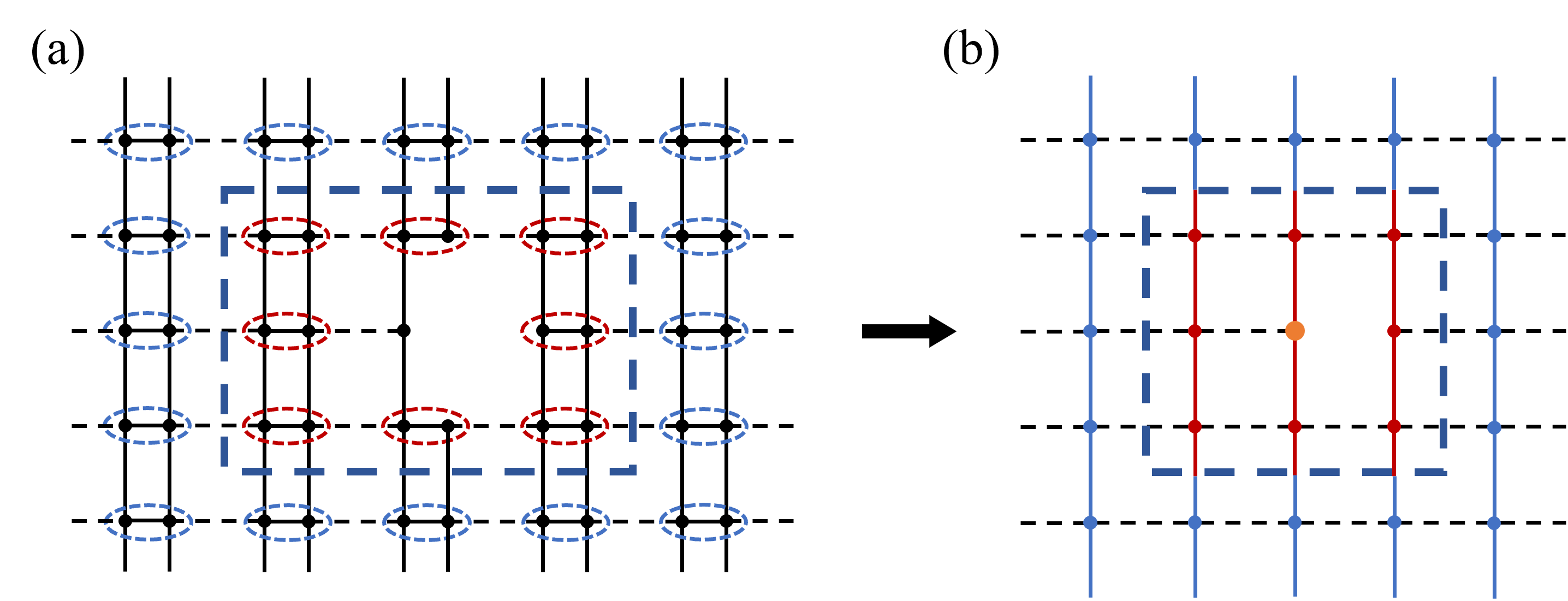}
\caption{\label{fig:old_ansatz}(a) The model defined on the coupled ladders with a defect, and the blue dashed rectangle denotes the impurity region, (b) the corresponding tensor-network representation. The dashed rectangle encloses the impurity tensors. Each tensor (except for the vacancy site) contains two physical degrees of freedom.}
\end{figure}

\begin{figure}[htbp]
\centering
\includegraphics[width=0.7\textwidth]{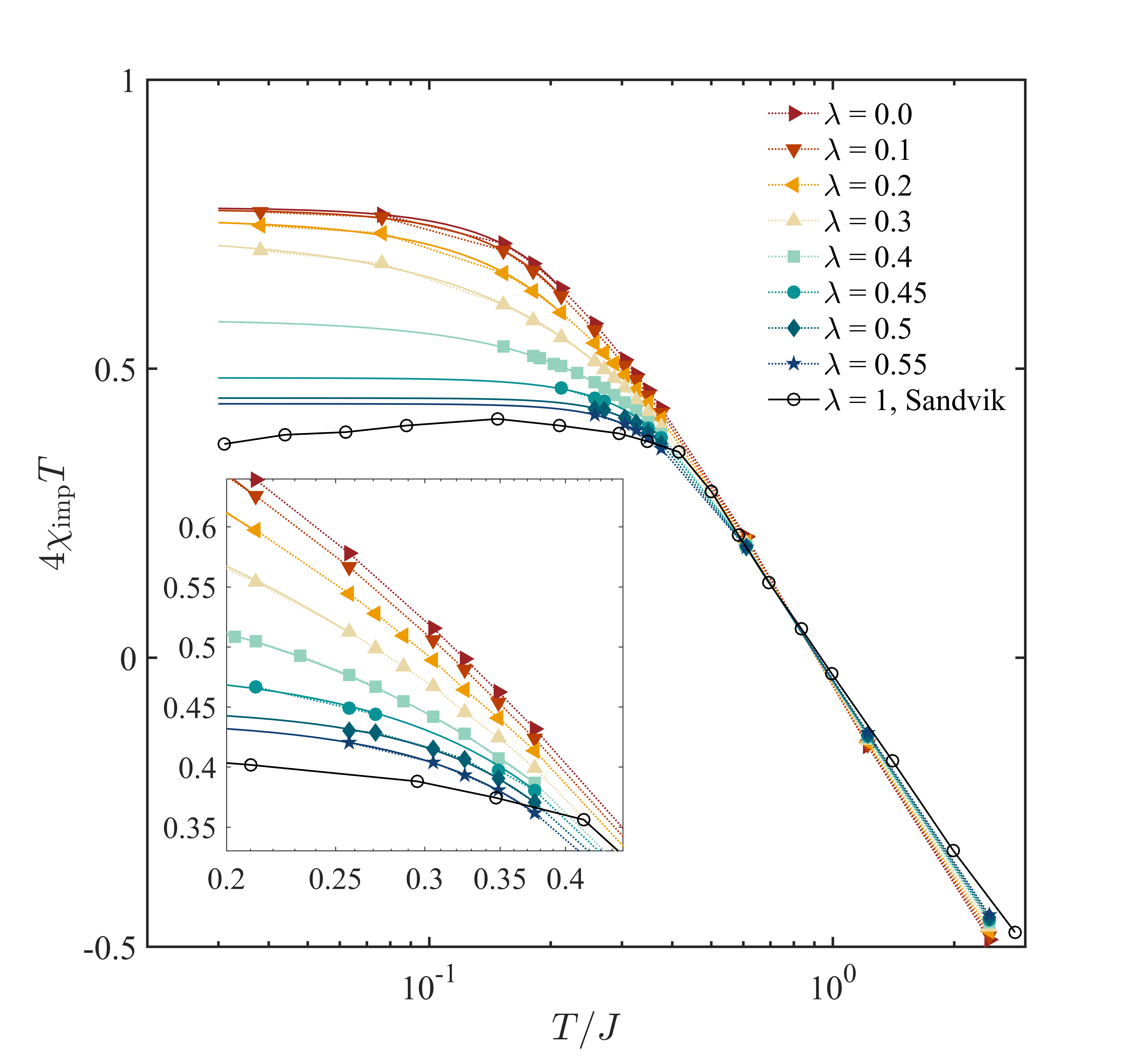}
\caption{\label{fig:old_result1} The impurity magnetic susceptibilities and its derivatives varying with $\lambda$.}
\end{figure}

\section{Dependence on the geometric shape of the impurity ansatz}\label{ansatz2}
\begin{figure}[htbp]
\centering
\includegraphics[width=0.7\textwidth]{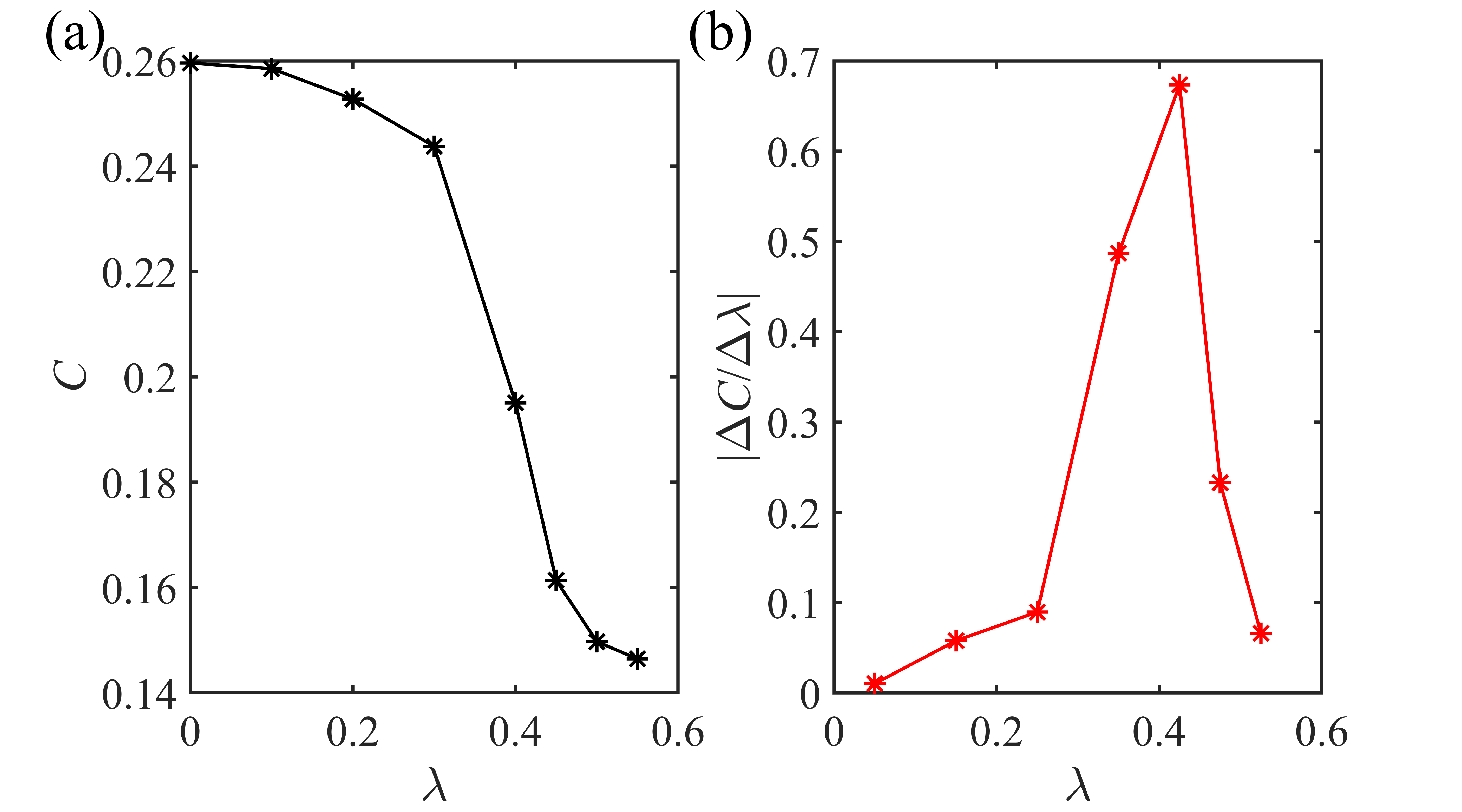}
\caption{\label{fig:result2}(a) Low-temperature plateau value $C$ extracted from $\chi_{\text{imp}}T$. (b) The derivative of $C$ with respect to $\lambda$.}
\end{figure}

As illustrated in Sec.~\ref{Sec:algorithm}, the IRG approach introduces a spatial boundary of the impurity propagation, thereby the obtained impurity properties would have a quantitative dependence on the size and shape of the impurity ansatzs. The dependence on the sizes has been demonstrated in the Heisenberg ladder in Sec.~\ref{Sec:benchmark}. In the following, we will demonstrate the dependence on the shape of the ansatz. As will be shown, despite slight quantitative shifts, all of our conclusions and results remain the same for different ansatz shapes. 

We employ a different impurity tensor-network ansatz to study the same coupled Heisenberg ladder model in the main text. The ansatz is shown in Fig.~\ref{fig:old_ansatz}. Using the IRG method, we examine the Heisenberg ladder model with an impurity at different temperatures across a range of $\lambda$ values from 0 to 0.55, encompassing the quantum critical regime from the paramagnetic VBS phase to the N\'eel AFM phase.

For different $\lambda$ values and temperatures, the calculated $\chi_{\text{imp}}$ are shown in Fig.~\ref{fig:old_result1}. For $\lambda < 0.3$ and at low temperatures, $\chi_{\text{imp}}$ is of the form $\chi_{\text{imp}} \sim C/T$. For $\lambda  > 0.3$, achieving comparable accuracy requires significantly larger bond dimensions—particularly $D_v$ along the ladders—which greatly increases the computational cost. Consequently, reliable data is limited at low temperatures for $\lambda \ge 0.4$. Despite this, we still observe a tendency of saturation in $\chi_\text{imp}T$ at the lowest accessible temperatures. We therefore perform power-law extrapolations based on the available data (Fig.~\ref{fig:old_result1}). The QMC data (for $\lambda=1$) extracted from the work of Sandvik et al. are also shown for comparison \cite{sandvik2003}.

From these extrapolated results, we find that the temperature where $\chi_\text{imp}T$ starts to saturate shifts to higher $T$ with increasing $\lambda$. Meanwhile, the extrapolated plateau values $C(\lambda)$ decreases with $\lambda$, as displayed in Fig.~\ref{fig:result2}(a). Moreover, the decreasing rate changes non-monotonically, as shown by the derivative $\Delta C/\Delta \lambda$ in Fig.~\ref{fig:result2}(b). The derivative also exhibits a cusp, and the jump behavior starts around the critical point, in qualitative consistency with the results shown in the main text. 

Notably, the quantitative plateau values in Fig.~\ref{fig:result2}(b) have deviations from those reported in the main text. This discrepancy can be understood by examining the bond correlations. As analyzed in the main text, the variation of bond correlations shows different characteristics for different phases. For the VBS phase, the bond correlations are primarily confined within the ladders. Upon increasing $\lambda$ into the N\'{e}el phase, they diffuse to the inter-ladder bonds, while simultaneously fading along the legs. Consequently, the ansatz introduced in this section, which retains additional sites on neighboring ladders around the vacancy, could provide a more accurate description of the N\'eel phase. This improvement is clearly visible in Fig.~\ref{fig:old_result1}. Compared to Fig. 4(a) of the main text, the impurity susceptibility on N\'eel side is in better agreement with the results of Ref[58] in the main text reference list. However, the ansatz here has its own downsides. The fewer impurity tensors retained on the ladder hosting the vacancy leads to a poorer description of the VBS phase and a deviation of the transition point. Nevertheless, it is clear from the above comparison that, although there are quantitative shifts of results, the quantitative impurity behavior remains the same for different ansatz choices.

\begin{figure}[htbp]
\centering
\includegraphics[width=0.7\textwidth]{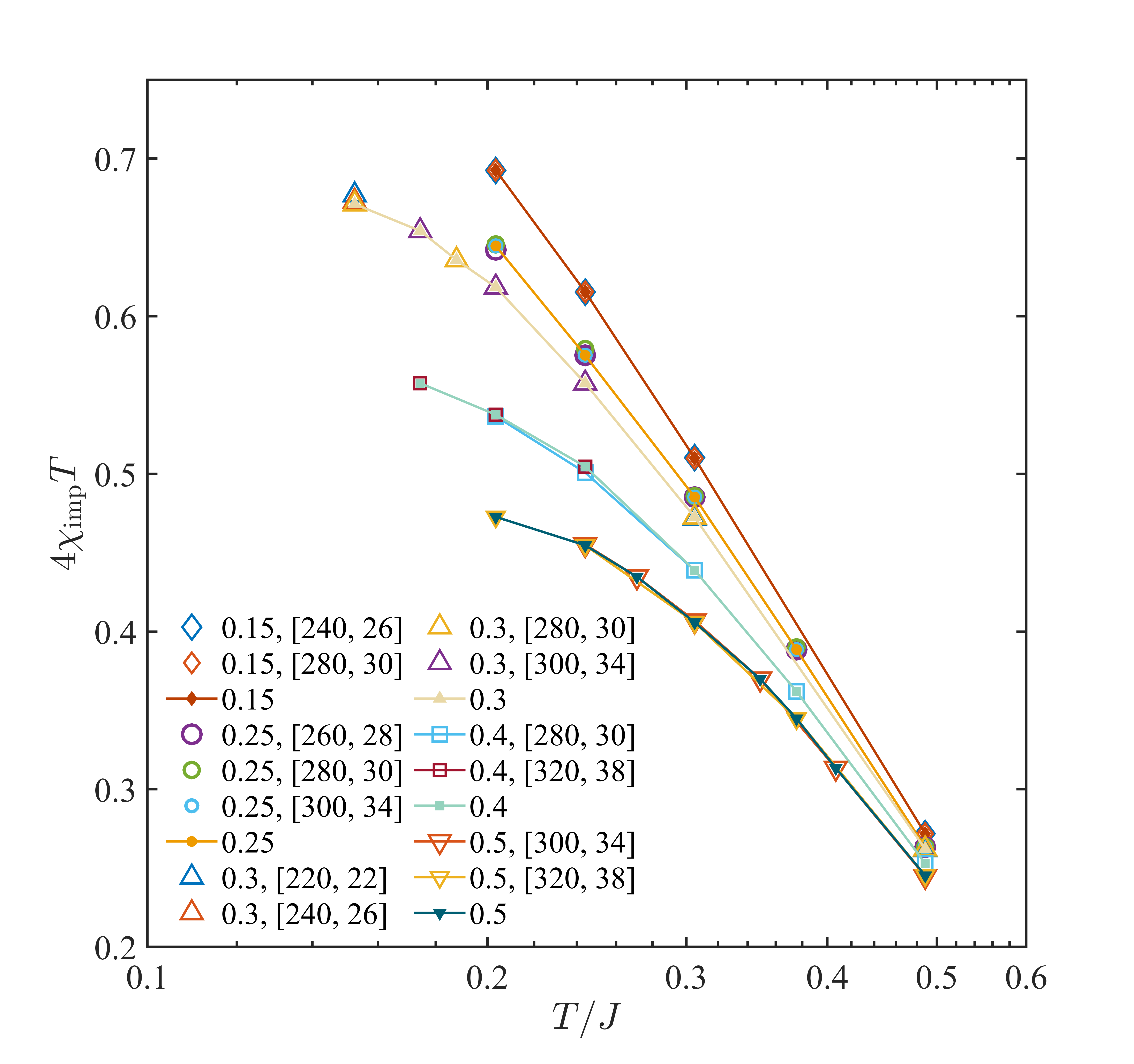}
\caption{\label{fig:dimension_choosing}Calculated $\chi_{\text{imp}}$ for various $\lambda$ and bond dimensions. The hollow markers indicate results corresponding to bond dimension configurations denoted by $\lambda, [D_v, D_h]$, where $D_v$ and $D_h$ represent the vertical and horizontal bond dimensions, respectively, and the solid markers denote the chosen results used in the analysis of the main text.}
\end{figure}

\section{Convergence of numerics}
The impurity susceptibility $\chi_\text{imp}$ is a crucial quantity derived from the magnetization of both bulk and impurity tensors. The precision of $\chi_\text{imp}$ relies heavily on the computational accuracy in terms of both the bulk and impurity components. In addition, the larger $\lambda$ requires larger bond dimensions to ensure sufficient accuracy, which significantly increases the computational cost. Thus, the convergence needs to be confirmed carefully, which will be demonstrated in the following. 

As illustrated in Fig.~\ref{fig:dimension_choosing}, we present a detailed convergence test where hollow markers represent the raw data obtained from calculations with increasingly larger bond dimensions used in LTRG, while the solid markers highlight the converged results selected for our final analysis in the main text.  Fig.~\ref{fig:dimension_choosing} guarantees that all the conclusions in the main text are based on well-converged numerical data, minimizing potential artifacts arising from insufficient bond dimensions. The convergence criterion is carefully chosen to balance the computational efficiency with the accuracy, ensuring that the selected bond dimension $\chi_\text{imp}$ reliably captures the impurity states. The convergence of the data used in Fig.~\ref{fig:old_result1} is confirmed following the same procedure.

\begin{figure}[htbp]
\centering
\includegraphics[width=0.4\textwidth]{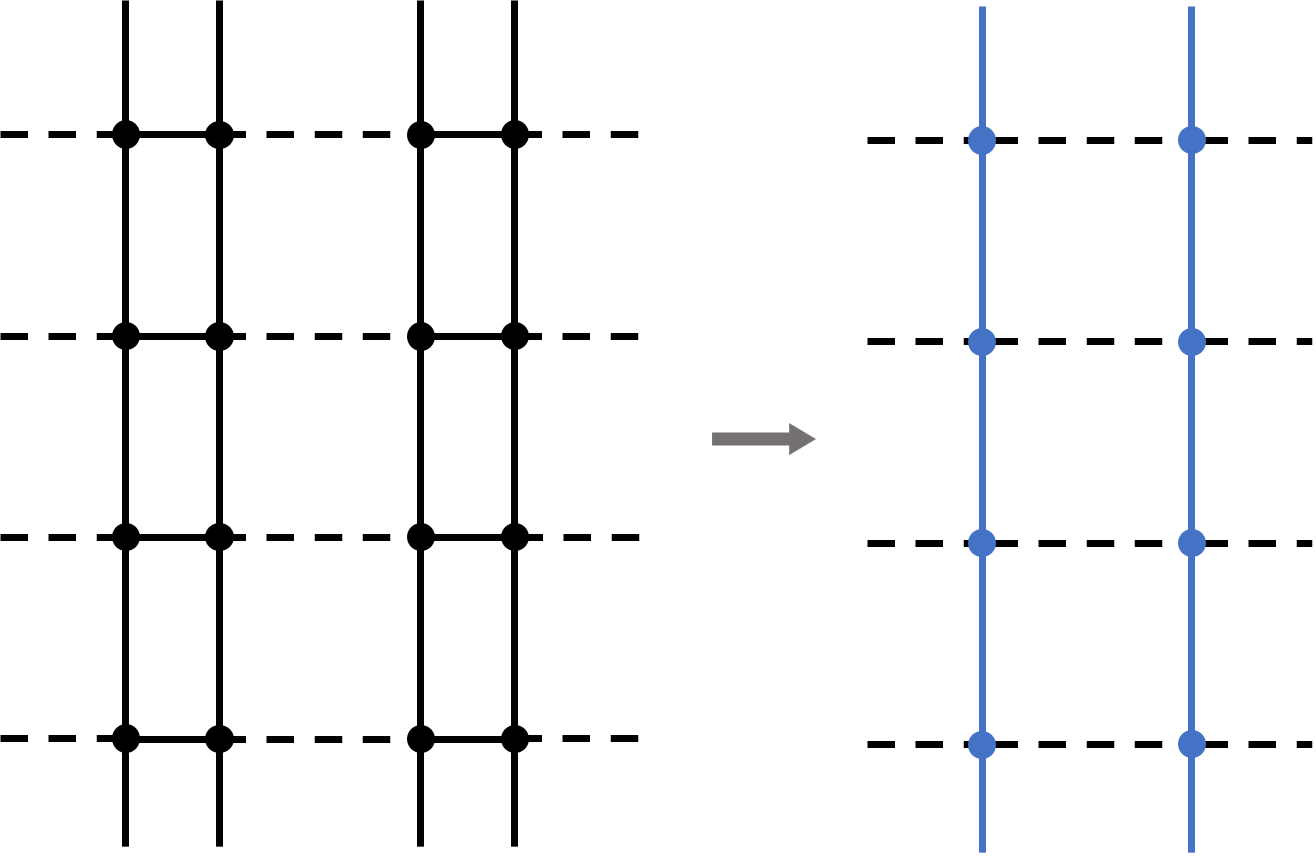}
\caption{\label{fig:bulk_ansatz}The tensor-network ansatz used to calculate the ground state of the coupled Heisenberg ladder model.}
\end{figure}
\section{Technical details on solving the ground state}

We employ tensor-network methods to solve the ground state of the coupled Heisenberg ladder model. The key step is to choose an appropriate tensor-network ansatz capable of capturing the essential entanglement of all possible phases. To this end, we design a specialized spatial projected entangled pair states (PEPS) ansatz tailored to the characteristics of the valence bond solid (VBS) state, which exhibits weak inter-ladder correlations. Specifically, we group two sites on the same rung as a cluster and regard their local Hilbert space as a single physical index, as illustrated in Fig.~\ref{fig:bulk_ansatz}. Compared to the conventional ansatzs, such as the projected entangled simplex states (PESS) and the conventional PEPS, this ansatz exhibits a superior performance in capturing the magnetization behavior of the studied model, which serves as a key signature in the underlying quantum phase transition. The ground states are obtained via the imaginary-time evolution, facilitated by the simple update method \cite{HCJiang2008}. By examining both sides of the critical point $\lambda_c$, we observe that the ground state magnetization remains close to zero for $\lambda \leq 0.2$, while it becomes finite for $\lambda \geq 0.3$, as displayed in Table \ref{magnetization}, in accordance with the phase diagram induced by energetics in the main text.

\begin{table}[htbp]
\centering
\begin{tabular}
{l|c|r}
\hline
$\lambda$   & Bond Dimensions & Magnetization \\ \hline
0   & 16, 2          & $5.19 \times 10^{-12}$       \\
0.1 & 16, 4          & $6.75 \times 10^{-12}$       \\
0.2 & 16, 4          & $5.12 \times 10^{-3}$        \\
0.3 & 16, 5          & 0.153                  \\
0.4 & 16, 5          & 0.296        \\ \hline
\end{tabular}
\caption{\label{magnetization}Magnetization of ground states.}
\end{table}